\documentclass[twocolumn]{aastex61}
\usepackage{CJKutf8}
\usepackage[T1]{fontenc}
\usepackage{alltt}
\usepackage{amsmath}

\hypersetup{linkcolor=blue,citecolor=blue,filecolor=cyan,urlcolor=magenta}
\submitjournal{ApJ}

\shorttitle{A momentum conserving $N$-body scheme}
\shortauthors{Q. Zhu}
\newcommand{\ON}{$\mathcal O({\textit N})$}

\newcommand{\rev}[1]{{\color{black}{{#1}}}}

\begin{document}
\begin{CJK}{UTF8}{gbsn}
\title{A momentum conserving $N$-body scheme with individual timesteps}
\author{Qirong Zhu(朱七荣)}
\email{Email: qxz125@psu.edu, zhuqirong1874@gmail.com}
\affiliation{Department of Astronomy \& Astrophysics; Institute for Cosmology and Gravity, The Pennsylvania State University, PA 16802, USA}
\affiliation{Harvard-Smithsonian Center for Astrophysics, Harvard University, 60 Garden Street, Cambridge, MA 02138, USA}

\begin{abstract}
$N$-body simulations study the dynamics of $N$ particles under
the influence of mutual long-distant forces such as gravity. 
In practice, $N$-body codes will violate Newton's third law if they
use either an approximate Poisson solver or individual timesteps. 
In this study, we construct a novel $N$-body scheme by combining 
a fast multipole method (FMM) based Poisson solver and a time integrator 
using a hierarchical Hamiltonian splitting (HHS) technique. 
We test our implementation for collision-less systems
using several problems in galactic dynamics. 
As a result of the momentum conserving nature of these two key components, 
the new $N$-body scheme is also momentum conserving. Moreover, we can 
fully utilize the $\mathcal O(\textit N)$ complexity of FMM with the integrator. 
With the restored force symmetry, we can improve both angular momentum 
conservation and energy conservation substantially. The new scheme will be 
suitable for many applications in galactic dynamics and structure formation. 
Our implementation, in the code {\sc Taichi}\footnote{ 
\url{https://bitbucket.org/qirong_zhu/taichi_public/}. The name ``Taichi(太极)" 
is inspired by the symmetries in the force computation and the time integration. 
The name also refers to the recursive relations and the hierarchical nature of 
FMM and the time integrator.}, is publicly available.
\end{abstract}



\section{Introduction} \label{sec:intro}
The conservation of momentum is a manifestation of Newton's third
law. However, current $N$-body codes do not conserve momentum
by introducing spurious net velocities, a direct consequence that
most of the Poisson solvers are not momentum conserving and individual
timesteps are not momentum conserving either. To some extent, 
momentum conserving error reveals the overall accuracy of $N$-body 
simulations. 

As an approximate Poisson solver, the most famous BH tree 
method \citep{Barnes1986} dramatically speeds up force evaluations 
by organizing particles into oct-trees. To any sink particles, 
source particles within a distance cell can be efficiently 
approximated by a single particle. However, there is no guarantee 
that the approximation is symmetric when the role of sink and source
reversed. Therefore, Newton's third law is violated in the force evaluation phase. 

For the time integration, adaptive individual timesteps play a crucial role 
in $N$-body codes by reducing the total number of force evaluations. 
However, force symmetry is broken between two particles with different 
timesteps. For the popular kick-drift-kick (KDK) integrator, particles with
smaller timesteps receive kicks at a higher frequency than particles with
larger timesteps while the positions of particles are not synchronized.
\rev{While many modern codes use synchronized positions for inactive
particles based on the velocities, this extrapolated position is slightly 
offset from the positions to perfectly respect the force symmetries 
with the active particles.} As a result, 
\rev{the exact} momentum conservation is also broken since the total 
kicks on all the particles do not cancel out. Again, we violate the force 
symmetry in Newton's third law.

In this study, we propose a new momentum-conserving $N$-body 
scheme by combining two momentum-conserving components: 
\rev{a Poisson solver based on the fast multipole method (FMM hereafter)}
and a timestep integrator \rev{based on a hierarchical Hamiltonian splitting 
(HHS hereafter)} proposed by \cite{Pelupessy2012}. 
We note that neither of the two components is completely new to the community. 
Examples of FMM codes include {\sc HOT} \citep{Warren1995}, 
{\sc falcON} \citep{Dehnen2000, Dehnen2002}, and {\sc PKDGRAV3} 
\citep{Potter2017}. FMM is also implemented as 
a short-range force solver in {\sc SWIFT} \citep{Schaller2016}, 
an ongoing project. The use of momentum conserving 
integrator can be traced back to the studies of planetary dynamics in the 
90s \citep[e.g.,][]{Wisdom1991, Duncan1998}. The {\sc HUAYNO} code by 
\cite{Pelupessy2012} includes the implementation of the HHS integrator. 
Recently, a TreePM code {\sc Arepo} \citep{Springel2010} 
with an implementation of the HHS integrator has been used in 
large scale cosmological production simulations \citep[see][]{Springel2017}. 

\rev{As we show later, the proposed new $N$-body scheme 
has some attractive features which might interest the community.
In particular, this scheme offers substantial improvements
in its accuracy over the existing practices.}
In \S~\ref{sec:method}, we briefly review the Poisson solver and time
integrator and outline our implementations. 
In \S~\ref{sec:result}, we show the performance of
the new scheme using a cold collapse test, an isolated disk 
galaxy, \rev{and the growth of a Milky Way-sized halo in 
a cosmological setup}. We further discuss the prospects 
of this new scheme in \S~\ref{sec:discussion}. 
We conclude with a summary in \S~\ref{sec:summary}. 

\section{Method} \label{sec:method}
\subsection{A FMM based Poisson solver}
The tree method by \cite{Barnes1986} approximates the gravitational force 
at a given sink particle for a group of source particles. In contrast, FMM 
starts with the expansion of gravitational potential at a group of sink 
particles in cell A with the cell centered at $\mathbf{z}_A$ and a group 
of source particles in cell B centered at $\mathbf{z}_B$. The key component 
of FMM maps the potential landscape generated by sources onto the sinks 
via a \textit{cell-cell} interaction. This cell-cell interaction is performed for 
all the cell pairs if they are well-separated according {\rev{some geometric 
or dynamic criteria. For example, we can use}} 
\begin{equation}
\lvert \mathbf{z}_A  -  \mathbf{z}_B \rvert > (R_A + R_B) / \theta, 
\label{eq:wellseperated}
\end{equation}
where $R_{A, B}$ is the cell size and the multipole-acceptance 
criterion (MAC) is denoted by a cell opening angle $\theta$. \rev{While 
other criteria could offer advantages in terms of speed or force accuracy 
\citep[e.g.,][]{Dehnen2002, Dehnen2014}, the simple geometric 
condition in Eq.~(\ref{eq:wellseperated}) is easy to implement.}

Besides the opening angle, the approximation errors are also 
controlled by the expansion order $p$. Unlike other Poisson solvers, 
FMM does not suffer the errors associated 
with the mass assignment, potential interpolation or aliasing. One 
distinct feature of FMM is that there is no distinction between the 
source and sink cells by construction. Therefore, this gravity solver is 
manifestly momentum conserving because the mutual forces are 
symmetric \citep{Dehnen2000, Dehnen2002} between cell A and B.
Moreover, the force symmetry holds true for \textit{any pair} consisting
of particle $i$ from cell A and particle $j$ from cell B:
\begin{equation}
\mathbf{F}_{ij} = - \mathbf{F}_{ji}
\end{equation}
Naturally, the total force between all the particles in cell A and cell B 
also cancels out as $\mathbf{F}_{AB} = -\mathbf{F}_{BA}$. In addition, 
when a desired magnitude of error is specified, FMM has an \ON\ complexity 
because the higher order terms decay faster than the increase of total 
particle number $N$.

The original FMM method proposed by \cite{Greengard1987} 
groups all the particles into a uniform grid. The data structure of 
a uniform grid is not optimal for highly irregular distributions. 
Alternatively, the same oct-tree structure used in BH method has been 
introduced to FMM \citep{Warren1995, Dehnen2000}. An efficient 
dual tree walk is implemented by these two studies as well. 
Starting from the root cell, we test the cell from the source tree 
and the sink tree according to Eq.~(\ref{eq:wellseperated}). 
If cell A and B are not well-separated, we split the cell with larger size 
and process its eight child cells. If neither A and B contain any more child 
cells, a direct summation is carried out for all the particles in cell A and B. 

The combination of oct-tree and dual tree walk forms the basis of a 
versatile and adaptive Poisson solver. For a detailed mathematical 
description of FMM and the tree walk procedure, we refer the readers to 
\cite{Dehnen2000, Dehnen2002} and \cite{Dehnen2011}. 

\iffalse
The explicit expression of multipole expansion in Cartesian coordinates, 
as in {\sc PKDGRAV}, can be quite cumbersome for high order expansions
$(p>3)$. For comparison, multipole expansion in spherical coordinates 
using spherical harmonics\citep[][]{Dehnen2014} leads to a compact 
expression for multipole calculations and interactions for any $p$. Note that 
Cartesian expansion and spherical harmonics are entirely equivalent for 
the same $1/r$ potential.
\fi

\subsection{A HHS time integrator with individual timesteps}

As we explained in the introduction, adaptive individual timesteps
introduce errors in momentum conservation by directly violating 
Newton's third law. \cite{Pelupessy2012} proposed a time 
integrator with individual timesteps which is free from this error. 
With the notations from \cite{Pelupessy2012}, 
the Hamiltonian of an $N$-body 
system $\mathcal{H}$ consists of a momentum term $\mathcal{T}$ and 
a potential term $\mathcal{V}$ as
\begin{equation}
\mathcal{H} = \mathcal{T}  + \mathcal{V} = 
\sum_{i}^{N}  \frac{\mathbf{p}^2_i}{2m_i} - 
\sum_{i, j \in N,\, i < j} \frac{Gm_im_j}{\lvert \mathbf{r}_i-\mathbf{r}_j \rvert}. 
\end{equation}
The fundamental idea is to split $\mathcal{H}$  into a slow system $S$ with its own
Hamiltonian $\mathcal{H}_S$  and a fast system $F$ 
with  $\mathcal{H}_F$ with a pivot timestep as
\begin{equation}
\mathcal{H}     = \mathcal{H}_S + \mathcal{H}_F,
\label{eq:splitting}
\end{equation}
where \begin{equation}
\mathcal{H}_S =  \mathcal{T}_{S} + \mathcal{V}_{SS} + \mathcal{V}_{SF},
\label{eq:hs}
\end{equation}
and 
\begin{equation}
\mathcal{H}_F = \mathcal{T}_{F} + \mathcal{V}_{FF}.
\label{eq:hf}
\end{equation}
$\mathcal{H}_F$ itself does not 
contain any contribution from system $S$. Therefore, the evolution of 
$F$ can be cleanly separated from $S$. The mutual 
interactions between $F$ and $S$ in  
\begin{equation}
\mathcal{V}_{SF}  = - \sum_{i \in S,\,  j \in F} \frac{Gm_im_j}{\lvert \mathbf{r}_i-\mathbf{r}_j \rvert} 
\end{equation}
are only encountered at the timestep 
determined by system $S$ (referring to the HOLD integrator, 
\rev{with $\mathcal{V}_{SF}$ `held' on the slow timestep}), as in Eq.~(\ref{eq:hs}).
This Hamiltonian splitting is then applied to $F$ with half of the original 
pivot timestep on $F$, which leads to a new $F$+$S$ splitting. 
The splitting then proceeds in a hierarchy of power of two till no further 
splitting can be found. The following pseudocode describes such iteration 
in the HOLD integrator constructed with a second-order accurate KDK s
integrator. \rev{After splitting the particles into a $F$ and $S$ system, 
we compute the forces between  $F$ and $S$ on the same rung $k$ 
while evolving the fast system $F$ on rung $k+1$.}
 
\begin{alltt}
\footnotesize
\textbf{ALGORITHM 1} 
EVOLE_HOLD(rung, system, time, dt, calc_timestep)

  slow, fast = split_slow_fast(dt, system);

  if(fast.size > 0)
    EVOLE_HOLD(rung+1, fast, time, dt/2, 0);

  KDK(rung, slow, fast, time, dt);

  if(fast.size > 0)
    EVOLE_HOLD(rung+1, fast, time+dt/2, dt/2, 1);
\end{alltt}

The splittings in Eq.~(\ref{eq:hs}) and (\ref{eq:hf}) introduce two different kicks 
in each hierarchy:  kickS2F \rev{(kick from slow to fast) due to} 
the term $\mathcal{V}_{SF}$ and 
kickFS2S \rev{(kick from slow and fast to slow) due to} the 
terms $\mathcal{V}_{SS} + \mathcal{V}_{SF}$. 
To kick system $F$ with kickS2F, we only need to include the forces 
originated from $S$. To kick system $S$, gravity from both $F$ and $S$ 
systems are evaluated. Our implementation of HOLD integrator 
contains variations from \cite{Pelupessy2012}. The following pseudocode 
describes the sequence of kicks and drifts operations in a KDK step within 
the HOLD integrator used in this study. \rev{The force/potential computation 
is part of each KICK operation.}

\begin{alltt}
\footnotesize
\textbf{ALGORITHM 2} 
KDK(rung, slow, fast, time, dt)

  if(fast.size > 0)
    KICK(rung, fast, slow, dt/2, 1, 0); //\textbf{kickS2F}
  
  KICK(rung, slow, fast, dt/2, 0, 0);   //\textbf{kickFS2S}

  drift(rung, slow, time, dt); 

  KICK(rung, slow, fast, dt/2, 0, 1);   //\textbf{kickFS2S}

  if(fast.size > 0)
    KICK(rung, fast, slow, dt/2, 1, 1); //\textbf{kickS2F}
\end{alltt}

\begin{figure}[ht!]
\plotone{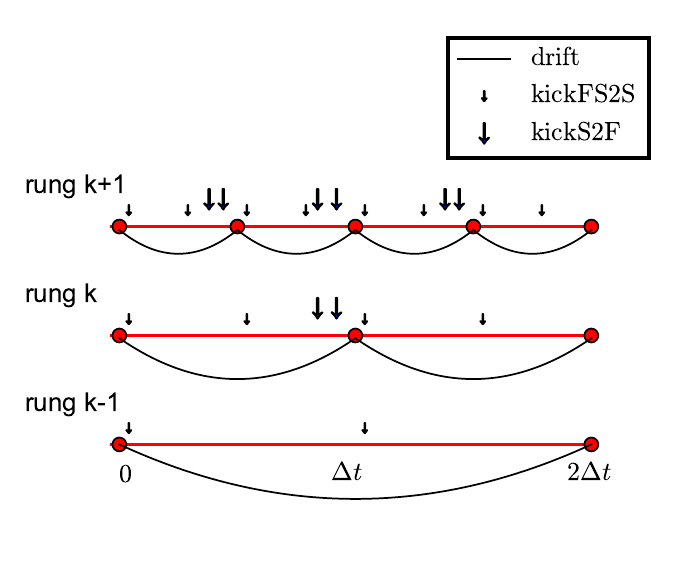}
\caption{A schematic of the HHS HOLD integrator. 
\rev{Here, particles on \rev{rung} $k$ march 
forward in time with a step size of $\Delta t$. 
In a normal KDK step, a drift operation advances 
the particle positions from the starting to the end 
points and two kick operations update their velocities 
at the start and in the middle of the timestep.}
The HOLD integrator injects two kickS2F, as indicated by large arrows, 
right before $t = \Delta t$. The frequency of kickS2F is determined by 
the step size of \rev{rung} $k-1$. Before each drift operation, 
the step size is adjusted by 
$\Delta t_{i} = \sqrt{\eta \epsilon /\mathbf{\lvert a \rvert}}$
by combining most recent kickS2F and kickFS2S.
\label{fig:hold}}
\end{figure}

Figure~\ref{fig:hold} illustrates the sequence of drifts and kicks
in the HOLD integrator. For particles on \rev{rung} $k$, we advance them
with a regular KDK integrator for a step size of $\Delta t$. Right
before $\Delta t$, two kickS2F operations from \rev{rung} $k-1$ are 
injected. It is clear that this integrator does not compute the total 
force on each particle.  Except for particles on the lowest \rev{rung}, 
their total forces are calculated by chance since there are no lower 
\rev{rungs} anymore. The individual timestep size is controlled by 
$\Delta t_{i} = \sqrt{\eta \epsilon /\mathbf{\lvert a \rvert}}$ with $\eta=0.02$, 
where we keep track of the accelerations from the previous kickS2F
or kickFS2S to obtain a conservative \rev{estimate} of the combined acceleration 
$\mathbf{\lvert a \rvert} \approx \mathbf{\lvert a_{\rm old} \rvert} + 
\mathbf{\lvert a_{\rm current} \rvert}$. Here, $\mathbf{a}_{\rm old}$ refers to the 
most recent kick from the \rev{lower} \rev{rungs} and $\mathbf{a}_{\rm current}$ is 
the most recent kick from \rev{particle with the same and faster timesteps}. 
Therefore, we update $\mathbf{a}_{\rm old}$ only at 
the second kick in each KDK step. Taking particles on \rev{rung} $k$ again in 
Figure~\ref{fig:hold} for example, the second kickFS2S is followed by the 
second kickS2F right before $\Delta t$ and the second kickS2F followed 
by another second kickFS2S right before $\Delta t $. \rev{Unfortunately}, 
we cannot recycle the previous kicks as in the common $N$-body codes 
due to the possible changes of \rev{rungs}. Instead, we evaluate gravitational 
potential and force for each kickS2F/kickFS2S in a KDK integrator. 
\rev{Therefore, the new scheme involves up to four separate calls to the 
Poisson solver within a single KDK step. In contrast, a traditional KDK step can 
be converted into a leapfrog integration reducing the number of calls to the
Poisson solver to half. Considering a system which contains only two rungs, 
$0$ and $k$,  we need to call Poisson solver $(2 + 2 + 2^{k+1})$ times 
for each step of rung 0. On the other hand, we can simply use a shared 
timestep based on rung $k$ thus reduce the total number calls to 
Poisson solver to $2^{k}$. As a result, the new scheme can be quite 
inefficient for those systems (due to the fact that shared timestep can 
benefit from a leapfrog integrator), in particular, when $k$ is small.}

The following pseudocode of KICK operation illustrates our treatment of 
kickS2F and kickFS2S as well as timesteps. Similarly, as {\sc huayno}, 
we treat sink and source particles separately. When forces 
on $F$ system are \rev{needed}, we mask the masses of all particles in $F$
to be zero. Their positions and masses are passed to a Poisson
solver together with all particles from $S$. As a result, only forces 
generated by $S$ on $F$ are computed. For kickFS2S, 
we use all the particles in both $S$ and $F$ \rev{systems} such that the all the 
accelerations from particles with finer timesteps than $S$ are included.

\begin{alltt}
{\footnotesize
\textbf{ALGORITHM 3} 
KICK(rung, system1, system2, dt, kickFS, secondkick)

  joint_pos_mass = get_pos_mass(system1, system2);

  if(kickFS)
    mask_mass_to_zero(system1);
  
  get_force_potential_from_exaFMM
    (joint_pos_mass, force, potential);

  set_acceleration_potentail(system1);

  update_velocities(system1, dt);

  if(secondkick)
    update_acceleration_old(system1); 
}
\end{alltt}

\subsection{Combining the two momentum conserving components}

\begin{figure}[ht!]
\plotone{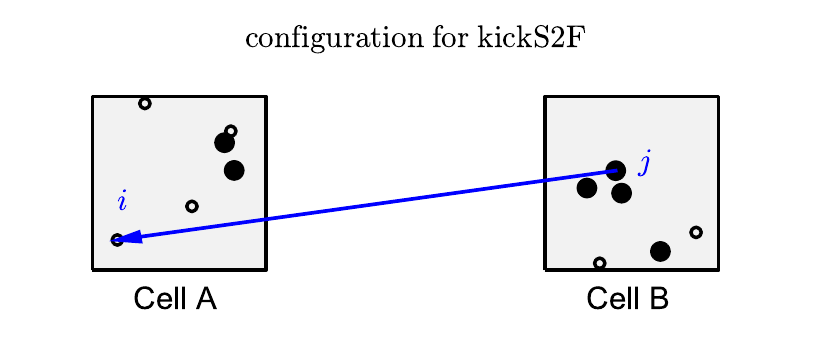}
\vspace{-0.25cm}
\plotone{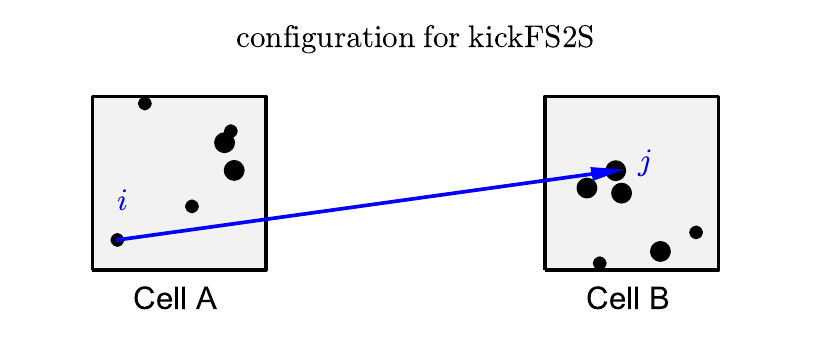}
\caption{Mass configuration of the slow ($S$, in the large symbols) 
and fast system ($F$, in the small symbols) for kickS2F and kickFS2S 
operations. The arrow represents the vector from a source particle 
to a sink particle. For kickFS2S, both the slow and (unmasked) fast system 
act as source particles. For kickS2F, we mask the 
mass of $F$ to be zero (in open circles) such that we only evaluate
the accelerations from $S$ to $F$. Force symmetry is present for 
any pair of particles drawn from cell A and B. 
Therefore, the HHS integrator with adaptive individual timesteps 
retains the force symmetry. 
\label{fig:mass_mask}}
\end{figure}

To implement the above ideas, 
we use {\sc ExaFMM}\footnote{\url{https://github.com/exafmm/minimal}. 
This is a minimal version of {\sc ExaFMM} with only spherical harmonics 
implemented for a Laplacian potential.}, 
a C++ FMM code, as our Poisson solver. 
\rev{{\sc ExaFMM} builds an oct-tree structure using a top-down approach 
based on recursion, with each cell in the oct-tree containing a particle 
number no less than $n_{\rm crit}$, which is set to be 64 in this study.}
\rev{{\sc ExaFMM} only includes a unsoftened Poisson solver, therefore
we introduce force softening with a Wendland $C^2$ function 
\citep{Wendland1995} by modifying the $1/r$ potential to}
\begin{equation}
\phi(r)=\frac{1}{\epsilon}(uu((u(u(3u-15)+28)-21)uu+7)-3), 
\end{equation} where $u = r/\epsilon$ for $r<\epsilon$. 
Since the spherical expansion does not apply to the softened potential, 
we additionally require the two cells can only enter cell-cell 
interaction when their distance is larger than the softening length 
$\epsilon$. 

\rev{The way {\sc huayno}\footnote{{\sc huayno} is 
available at \url{www.amusecode.org}.} handles these two different kicks is straightforward, 
where a mass array consisting of both source and sink particles is passed 
into its direct summation solver.}
\rev{To ensure the momentum conservation with our new scheme,
we set up two different mass configurations for kickSF and kickFS2S
separately.}
The upper panel of Figure~\ref{fig:mass_mask} shows 
the masking of the mass of particles in $F$ (in the open circles). Only forces
from $S$ (in the large solid circles) are responsible for generating the
gravitational potential for each particle in $F$. The lower panel 
shows the mass configurations for kickFS2S.
Due to the built-in symmetry in the multipole expansion, it is guaranteed 
that the force $\mathbf{F}_{ji}$ from particle $j$ in the top panel of 
Figure~\ref{fig:mass_mask} balances force $\mathbf{F}_{ij}$
in the lower panel. Therefore, momentum is conserved even we use 
two different mass configurations in kickS2F and kickFS2S separately. 
\rev{Of course, we must ensure the tree walkings in the separate
calls of kickS2F and kickFS2S are identical.}

Because {\sc ExaFMM} uses the geometric size (\rev{cell side length}) 
of each cell to test the well-separateness condition in 
Eq.~(\ref{eq:wellseperated}), we will encounter a situation where we 
can split either cell A \rev{or} B if they have equal 
size. This degeneracy is quite a subtle point that ruined the perfect force 
symmetry during tree traversal in our earlier attempts. To remove this 
asymmetry, we further require both cells to be opened if they are not 
well separated. \rev{Note this approach is suboptimal, a radius 
based on the smallest enclosing sphere \citep{Dehnen2014} do not 
lead to this degeneracy in the first place.} The following code is our dual 
tree walk procedure. {\sc ExaFMM} generates an interaction list for 
cell-cell interactions and a list for direct summation first. Then 
{\sc ExaFMM} evaluates the potentials and forces altogether 
looping through the interaction lists for each cell in its evaluations phase. 

\begin{alltt}
{\footnotesize
\textbf{ALGORITHM 4} 
GET_INTERACTION_LIST(Cell Ci, Cell Cj)

  if(well_seperated(Ci, Ci))
    Ci.M2L_list_add(Cj);

  else if(Ci == leaf and Cj == leaf)
    Ci.P2P_list_add(Cj);

  else if(Ci == leaf and Cj != leaf)
    for(all child cells cj of Cj)
      GET_INTERACTION_LIST(Ci, cj);

  else if(Ci != leaf and Cj == leaf)
    for(all child cells ci of Ci)
      GET_INTERACTION_LIST(ci, Cj);
    
  else
    for(all child cell pairs \{ci, cj\} of \{Ci, Cj\})
      GET_INTERACTION_LIST(ci, cj);
}
\end{alltt}

In each kick, we pass the position and mass arrays to {\sc ExaFMM} 
to complete a complete FMM evaluation step. The gravitational potentials 
and forces computed by FMM are then passed to the integrator to advance 
the system to the next step. The expense of each tree building is non-negligible. 
However, the hierarchical splitting of the total system dramatically 
reduces the cost of the average FMM computation. In each hierarchy, 
only a subset of particles participates in FMM as all the other particles 
in its lower hierarchies do not. \rev{As a result, the problem size of FMM 
is progressively reduced for the more dynamic part of the system. This 
feature is attractive for a system with a small fraction of very active particles. 
Even though we only need to evaluate the forces on the active particles, 
we still have to build the tree and evaluate the moments for all the particles, 
which can pose a significant overhead thus slows down the entire simulation.}
In a simplified case where each hierarchy is occupied with a 
smaller particles numbers (say half of the previous hierarchy), 
only an additional factor ($\sim \sum_{l=1}^{n}{1/2^{l}}$) is 
introduced in a total of $n\sim\ln(N)$ steps. As a result, the 
HHS time integrator can maintain the \ON\ complexity of FMM 
using individual timesteps. Therefore, the combination of these two key 
components offers two highly \rev{attractive} features into a new $N$-body 
scheme: momentum conserving and an \ON\ complexity. 

Our implementation is in the code {\sc Taichi}, which \rev{is largely} 
built upon on {\sc huayno} \rev{using its routines for Hamiltonian 
splitting} and {\sc ExaFMM} for \rev{force calculations}. 
\rev{The momentum conserving nature of our new scheme}
is demonstrated in our tests, as we show in the next sections. 
The fiducial values are \{$\theta$ = 0.45, p = 5\} if not stated otherwise.
Since we have used two momentum conserving components in 
{\sc Taichi}, we also implement a traditional integrator with individual 
timesteps with FMM as the Poisson solver. 
\rev{For each active step, we evaluate gravitational forces on all the 
active particles due to all the mass in the system. For the inactive 
particles, we only use their positions without
synchronization to the current timestep.}

\section{Results} \label{sec:result}

\subsection{Run-time efficiency of the new scheme}
\begin{figure}[ht!]
\plotone{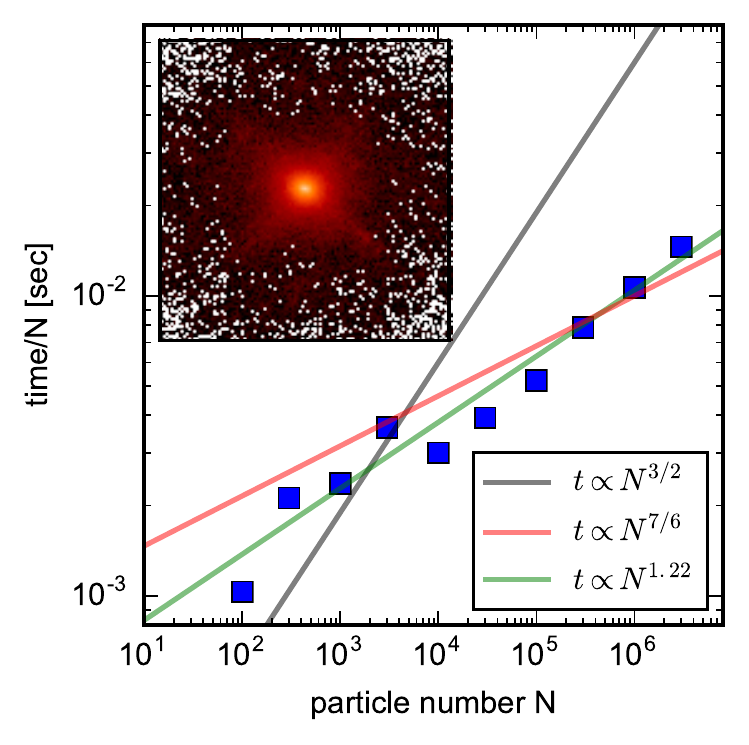}
\caption{Wall-clock time of a series of cold collapse simulation of 
particles from an initial cubic configuration with {\sc Taichi} code. 
The image shows the distribution of $3\times10^6$ particles at $t=1.0$ 
within the unit box. The system undergoes collapse at $t\approx0.5$
and forms a core-envelope structure. The measured wall-time scales 
\rev{as $N^{1.22}$, slightly above the rate of $N^{7/6}$}. 
\label{fig:walltime}}
\end{figure}

We first test the performance of the new scheme using a 
cold collapse of particles released from a cubic configuration. 
We place random points at rest with a total 
mass of unity within a cube of unit size with the 
total number varied from 100 to $3\times10^6$. We use H\'enon 
units with $G = M = R = 1$ and evolve the system from $t = 0$ to 1.
Besides, we vary the softening length $\epsilon$ according to 
$\epsilon=0.01(N/10^4)^{-1/3}$ to mimic the usual practice in the
community \citep{Power2003, Springel2008}. 
This simple setup is non-trivial for $N$-body codes as 
the system will undergo a quick collapse at $t\sim0.5$. As a 
result, a dense core forms in the center while a diffuse envelope 
due to ejected particles develops in the vicinity as shown in the image
in Figure~\ref{fig:walltime}. Throughout the 
simulation, the gravitational force varies strongly both in space 
and in time. Each simulation is performed in parallel using OpenMP 
with 16 cores on a quad 2.80GHz Xeon E5-2680 server node. 

Figure~\ref{fig:walltime} also shows the measured wall-clock time as a function 
of $N$ compared with two guiding lines of $N^{7/6}$ and $N^{3/2}$ trends. 
With the timestep function $\Delta t = \sqrt{\eta \epsilon/\lvert \mathbf{a} \rvert}$, 
one would expect the running time scale up between $N^{1/6}$ and $N^{1/2}$ 
due to the decreasing $\epsilon$ alone. The former bound is obtained assuming 
the typical acceleration does not change. The latter is obtained by 
considering the typical acceleration changes as $\epsilon^{-2}$ hence 
$(\Delta t)^{-1} \propto \epsilon^{-1.5}\propto N^{0.5}$. 
Both trends are assuming the \ON\ complexity for FMM force 
computation. Our test demonstrates that the performance
of the scheme agrees better with the $N^{7/6}$ trend than the other. 
\rev{Only with $N$ (<$10^4$), we see some collisional behavior 
described by $N^{3/2}$ due to the very small particle number.}
In addition, we compare the particle distribution in the dense core, 
which is indeed better resolved with higher resolutions. 
This result means the acceleration of a subset of particles is much stronger in higher 
resolutions. Therefore, the complexity of FMM perhaps is 
\rev{close to} \ON\ \cite[see also][]{Dehnen2002, Dehnen2014}. Nevertheless, 
we can retain the excellent efficiency of FMM in the new 
scheme even we invoke a full FMM procedure, including tree building,  
at every individual timestep.

\subsection{Momentum conservation, angular momentum conservation, 
and energy conservation of the new scheme}

\subsubsection{Exact momentum conservation}
Next, we examine the momentum conversation using the same collapse test. 
We set up the initial condition with a total number of $10^4$ and 
evolve it with {\sc Taichi}. We monitor the total momentum, angular 
momentum, and total energy during the simulation. 
At $ t = 1$, the \textit{r.m.s} velocity is close to unity. Therefore, 
we simply take the magnitude of the net momentum of the entire system 
$\lvert{\mathbf P}\rvert$ as a proxy for momentum conservation error. 

\begin{figure}[ht!]
\plotone{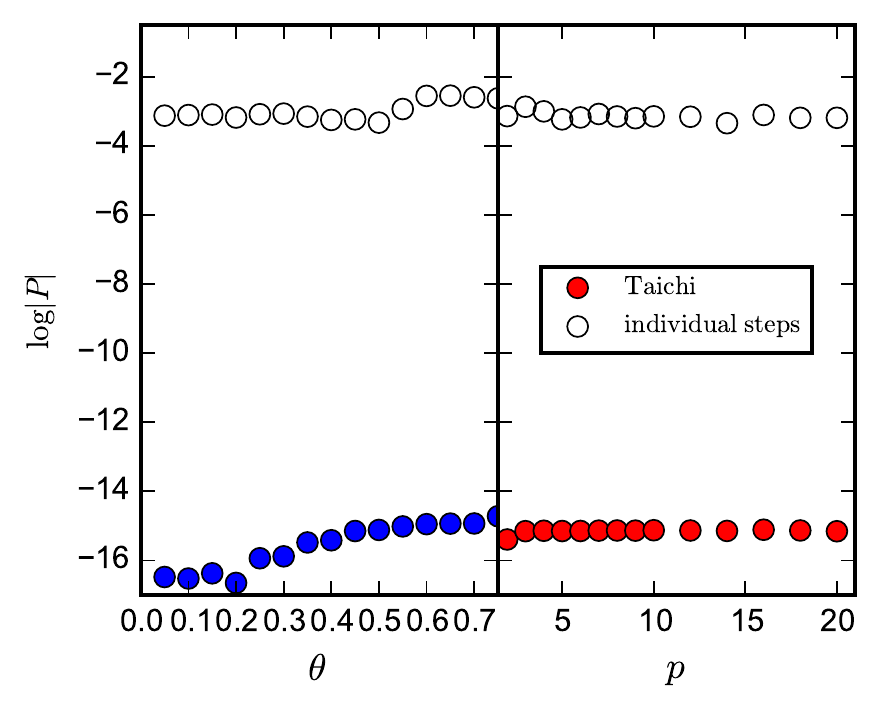}
\caption{Momentum conservation error at $t = 1$, as a function of $\theta$ 
and $p$, in the cold collapse test. We vary $\theta$ from 0.7 to 0.05 
while keeping $p=5$ fixed and $p$ from 3 to 20 with $\theta = 0.45$. The 
combination of FMM and HHS is momentum conserving while the error 
is at $\log(\lvert \mathbf{P} \rvert) \sim -3.2$ for the traditional integrator with 
individual timesteps. 
\label{fig:momentum_error}}
\end{figure}

Figure~\ref{fig:momentum_error} shows the momentum conservation error 
as a function of $\theta$ (left panel) and $p$ (right panel). The values of 
$\log(\lvert {\mathbf P} \rvert)$ confirm that our implantation is 
indeed momentum conserving, which validates our proposal in the 
previous section. There are some residual errors due to round-offs 
in FMM, which slowly increases with a larger opening angle $\theta$. 
\rev{The accumulation of the round-off errors depends on the inhomogeneity 
of the particle distribution and the parameter $n_{\rm crit}$. In practice, 
this tiny error should not be of any concern.}

In Figure~\ref{fig:momentum_error}, we also include the momentum conservation 
error with FMM but with the traditional integrator with individual 
timesteps in the open circles.  $\log(\lvert {\mathbf P} \rvert)$ stays rather 
flat at \rev{-3.2} with varying $\theta$ or $p$, which originates from the 
force asymmetries in the integration with individual timesteps. 

For tree codes, only in the joint limits of $\theta \rightarrow 0$ and 
using shared timesteps, we can effectively eliminate momentum 
conservation error. To verify this, we also run a tree code,
{\sc Gadget-2}\footnote{\url{https://wwwmpa.mpa-garching.mpg.de/gadget/}.}
\citep{Springel2001, Springel2005}, 
with the same initial condition. {\sc Gadget} uses two types of 
cell opening criteria: a standard geometric opening angle in BH 
tree codes and a relative criterion based on particle acceleration. 
With $\theta$ changes from 0.8 to 0.05, we find $\log \lvert \mathbf{P} \rvert$ 
reduced from -2.2 to -3.5. If we use the relative criterion
with force tolerance parameter $\alpha$ from 0.02 to $1\times10^{-4}$, 
$\log \lvert \mathbf{P} \rvert$ changes from -1.2 to -3.5\footnote{We find that
the relative criterion, although more efficient in tree walking given 
the same relative force error, introduced more momentum conserving error 
in this test. The net acceleration of the system is comparable between 
$\theta=0.6$--$0.7$ using the standard geometrical criterion of BH and 
$\alpha=1\times10^{-3}$ with the relative criterion.}. 
Because the monopoles in {\sc Gadget} can be updated with 
a half-step \citep{Springel2001, Springel2005} using the velocity
of the node, the error in momentum conservation is 0.5 dexes better
than FMM with individual timesteps in Figure~\ref{fig:momentum_error}. 
Nevertheless, the remaining error which does not converge away with 
smaller $\theta$, originated from individual timesteps.

\subsubsection{Angular momentum conservation}

\begin{figure}[ht!]
\plotone{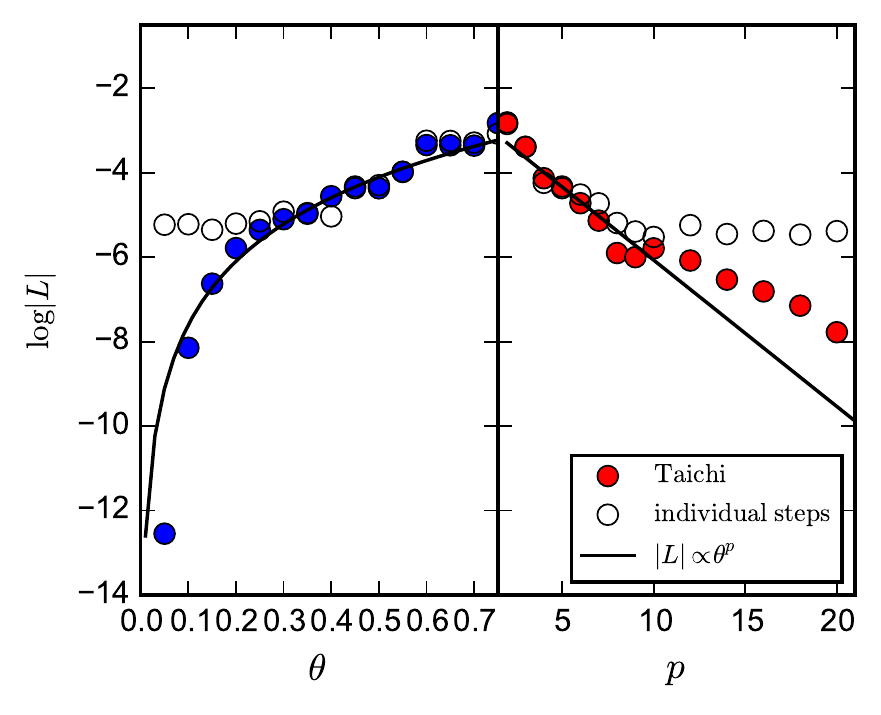}
\caption{Angular momentum conservation error at $t = 1$ vs. $\theta$ 
and $p$. The error decreases as a strong function with $\theta^{p}$. 
When $\theta\rightarrow0$, the error is reaching machine zero. Meanwhile, 
increasing the expansion order $p$ also reduces the angular momentum 
conservation error albeit at a slightly slower rate than with lowering $\theta$. 
With the traditional integration with individual timesteps, 
we cannot further reduce the error once $\theta < 0.25$ or 
$p > 7$ when the force asymmetry in the time integrator start to dominate 
over force asymmetry in space. 
\label{fig:angular_momentum_error}}
\end{figure}

While FMM conserves linear momentum by construction, 
it is well-known angular momentum is not conserved exactly 
due to the fact $\lvert \mathbf{F}_{ij} \times \mathbf{r}_{ij}\rvert \ne 0$. 
Recently, \cite{Marcello2017} has proposed a corrective torque term 
that restores angular momentum conservations on the cell level. 
(However, $\lvert \mathbf{F}_{ij} \times \mathbf{r}_{ij} \rvert \ne 0$ still holds 
true among particle pairs.) Since the force approximation error in FMM scales 
as $\theta^p$ \citep{Dehnen2002, Dehnen2014}, the angular momentum 
conservations error  also follows a similar trend. 

We use the net angular momentum $\lvert \mathbf{L} \rvert$ as a proxy 
for angular momentum conservation error. 
Figure~\ref{fig:angular_momentum_error} shows the angular 
momentum conservation error vs. $\theta$ and $p$.
Reassuringly enough, we can also reduce angular momentum conservation 
error with either a smaller $\theta$ or a larger $p$. In particular, 
a small $\theta$ can drastically reduce the error close to machine zero 
as in $\theta = 0.05$ case.

In contrast, a traditional integrator with individual timesteps gives a very 
different trend of $\log(\lvert \mathbf{L} \rvert)$ vs. $\theta$ and $p$. When the 
opening angle $\theta > 0.25$ or the expansion order $p < 7$, 
$\log(\lvert \mathbf{L} \rvert)$ shows a very similar trend as
{\sc Taichi}. On the other hand, 
we cannot further reduce $\log(\lvert\mathbf{L}\rvert)$ 
below \rev{-5.3} when the force asymmetry in 
the time integrator takes over the torque error in FMM in the total 
error budget. 

For comparison, $\log \lvert \mathbf{L} \rvert$ with {\sc Gadget} can be
reduced to $-5.4$ in its BH mode with $\theta=0.05$ or to $-6.0$ with
$\alpha=1\times10^{-4}$ with its relative criterion. These errors are 
slightly better than that obtained with FMM with individual timesteps but
significantly worse than {\sc Taichi}.

\subsubsection{Energy conservation}
\begin{figure}[ht!]
\plotone{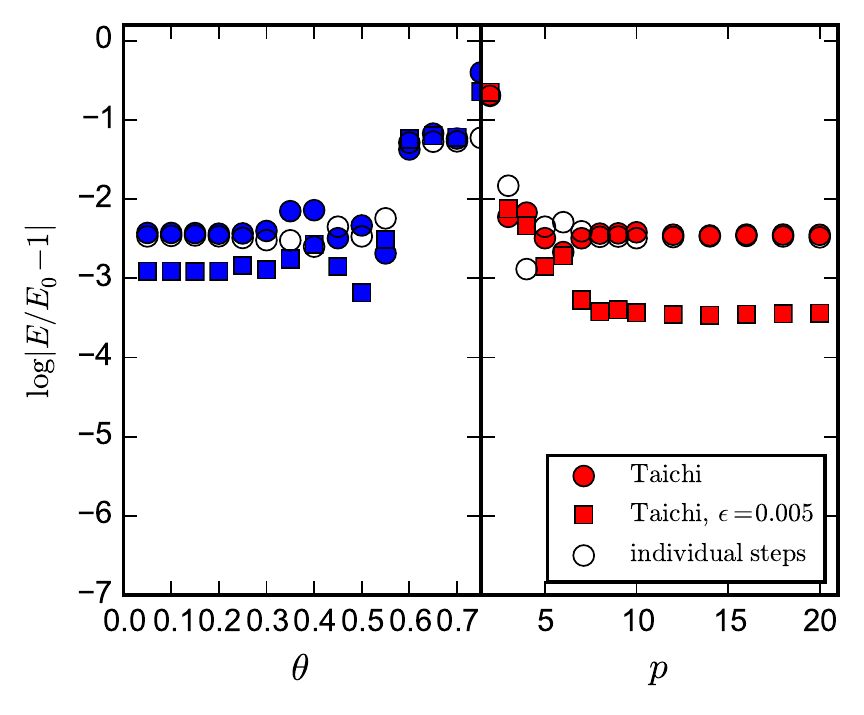}
\caption{Energy conservation error at $t = 1$, as a function of $\theta$ 
and $p$, in the cold collapse test. We also present {\sc Taichi} integrations 
with smaller timesteps using $\epsilon=0.005$. It is clear that force 
asymmetry in the traditional integration with individual timesteps is
also a source of error in energy conservation. 
\label{fig:energy_error}}
\end{figure}

The source of energy conservation error in $N$-body codes is a more 
complicated issue when compared with momentum and angular momentum.
In a fixed potential, energy conservation error can be reduced with smaller
step size or with a time-symmetric formulation of timestep choice 
\citep{Dehnen2011, Pelupessy2012, Dehnen2017}. Energy conservations 
errors due to time integration are extensively studied for test particles in a 
fixed potential \citep[e.g.,][]{Springel2005, Dehnen2017}.

Nevertheless, there are other sources of error in a time-varying potential, 
including the force asymmetry when integrated using the traditional individual 
timesteps. In Figure~\ref{fig:energy_error}, we compare the energy errors in terms of 
$\log(E/E_0 - 1)$ at $t = 1$, where $E$ is the total energy,  
with results obtained by {\sc Taichi} and the traditional individual timesteps. 
\rev{The energy error obtained with traditional individual timesteps
is very close the errors obtained with {\sc Taichi}.}
Additionally, we use a smaller timestep with $\epsilon = 0.005$ instead 
of 0.02 for {\sc Taichi}. As expected, the energy error can be reduced from -2.5 
to -3.2 with finer step sizes. Besides, the energy error does not show substantial 
systematic variation with $\theta$ or $p$. The energy conservation error 
flattens out with $\theta\rightarrow0$ or $p\rightarrow20$.

\subsection{The cost towards higher accuracy}
\begin{figure}[ht!]
\plotone{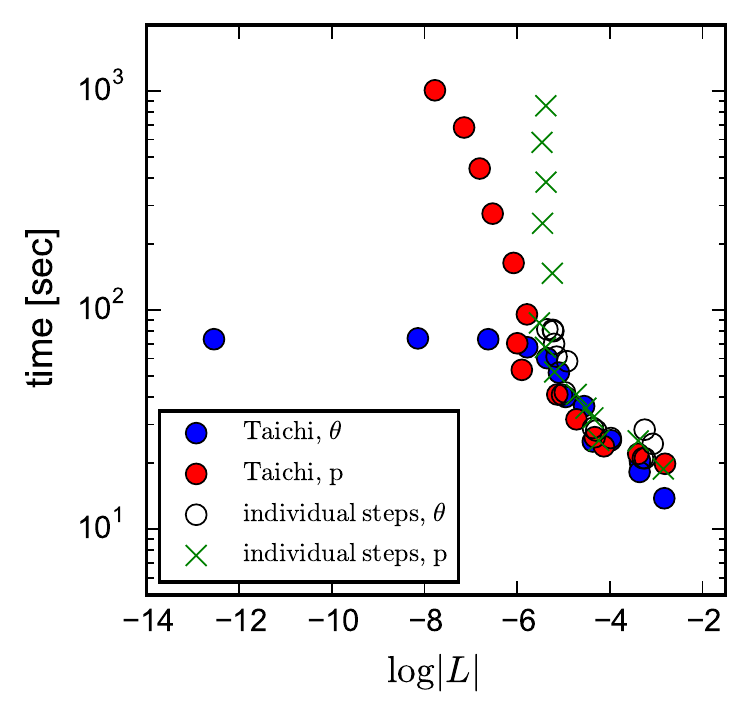}
\caption{The computational cost in terms of wall time vs. the error of total angular 
momentum $\log(\lvert \mathbf{L} \rvert)$ at $t = 1$. The errors decrease with 
$\theta\rightarrow0$ or with an increasing $p$, although the former strategy 
is more efficient given the same cost. On the other hand, angular momentum 
conservation error with individual timesteps can only be lowered to 
$\log(\lvert \mathbf{L} \rvert) \sim -6$ without any further improvement. 
\label{fig:error_cost}}
\end{figure}

With the momentum conserving nature of {\sc Taichi} by construction, 
and the energy conservation error from multiple sources, we use 
angular momentum $\log(\lvert \mathbf{L} \rvert)$ as a proxy for the 
overall accuracy of our code. In Figure~\ref{fig:error_cost}, we plot the 
computational cost in terms of wall time vs. the $\log(\lvert \mathbf{L} \rvert)$. 
Higher accuracies can be achieved in FMM with $\theta\rightarrow0$ 
or with an increasing expansion order.
It is more cost effective to use a smaller opening angle than a larger 
expansion order when higher accuracy is desired. On the other hand, 
the traditional integrator with individual timesteps cannot further 
improve once below $\log(\lvert \mathbf{L} \rvert) = -6$.

\begin{figure*}
\gridline{\fig{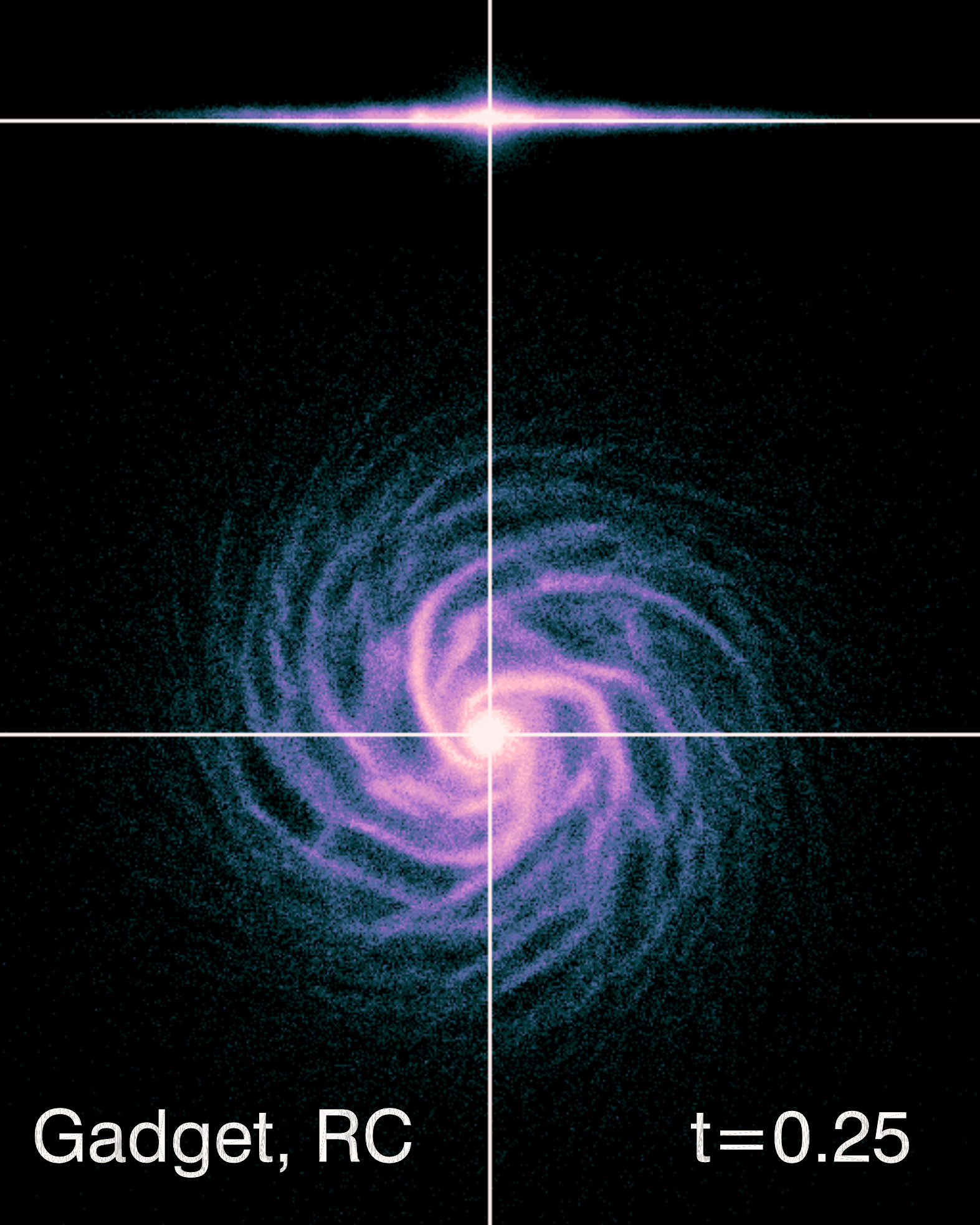}{0.27\textwidth}{}\
             \fig{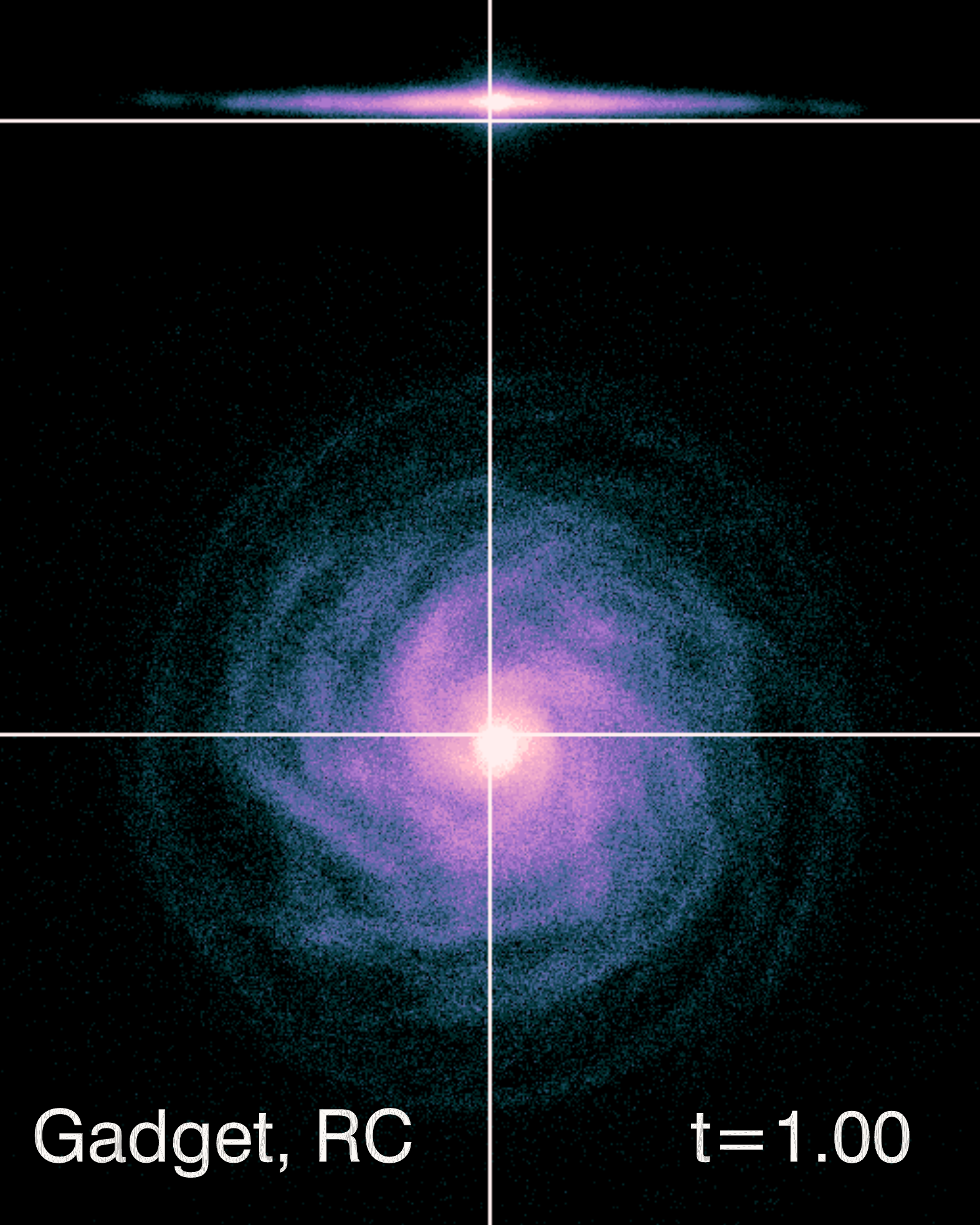}{0.27\textwidth}{}\
           \fig{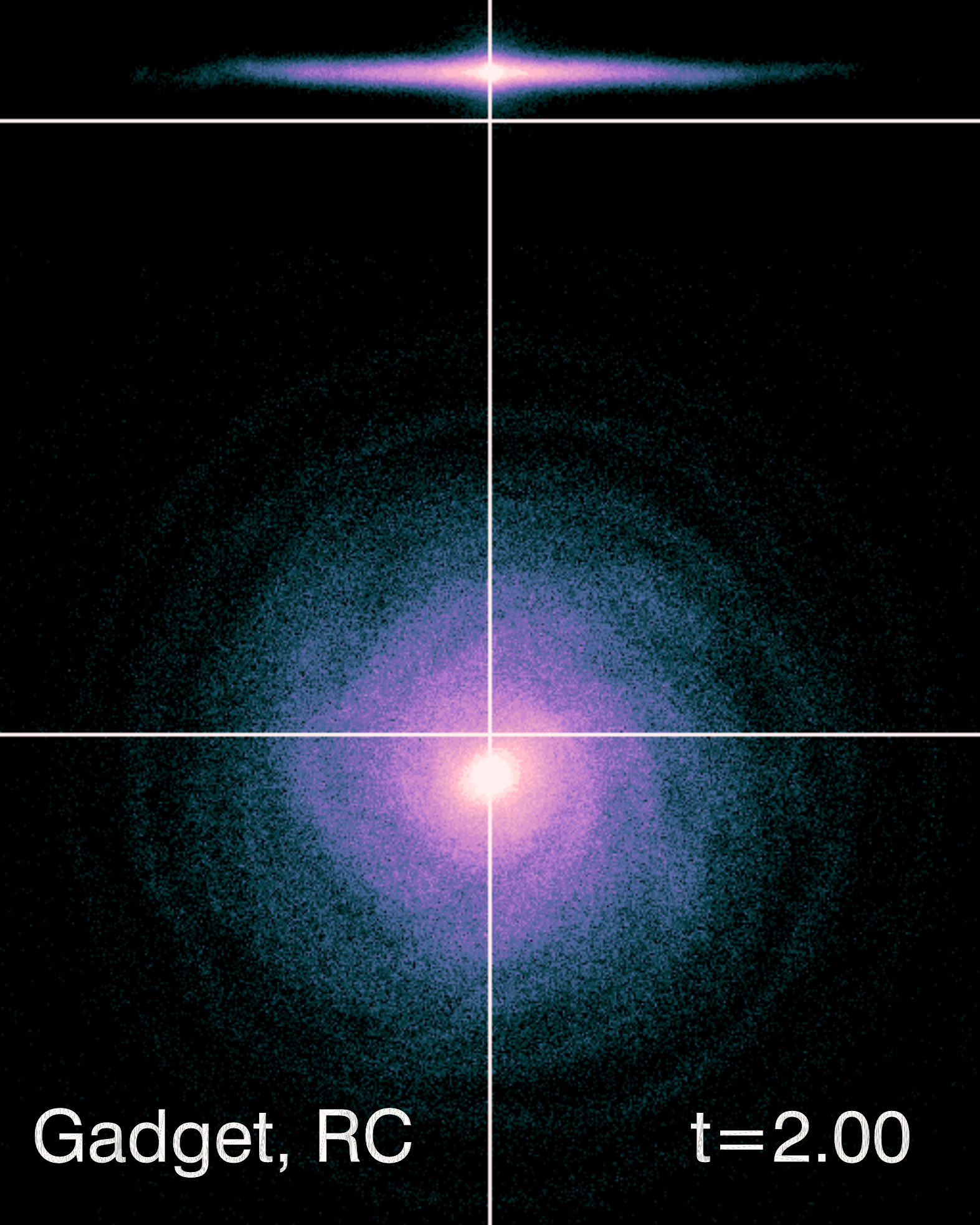}{0.27\textwidth}{}}
\gridline{\fig{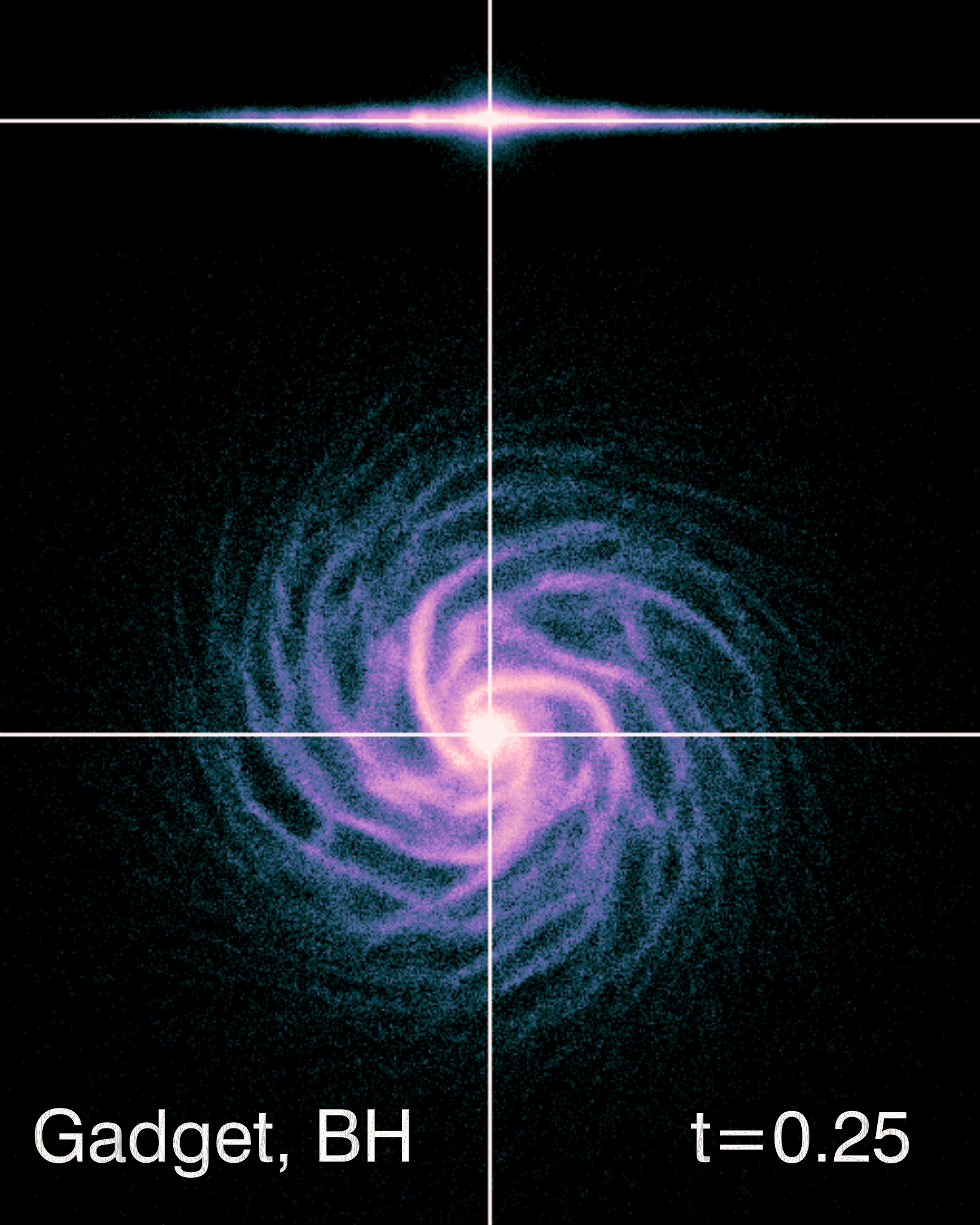}{0.27\textwidth}{}\
              \fig{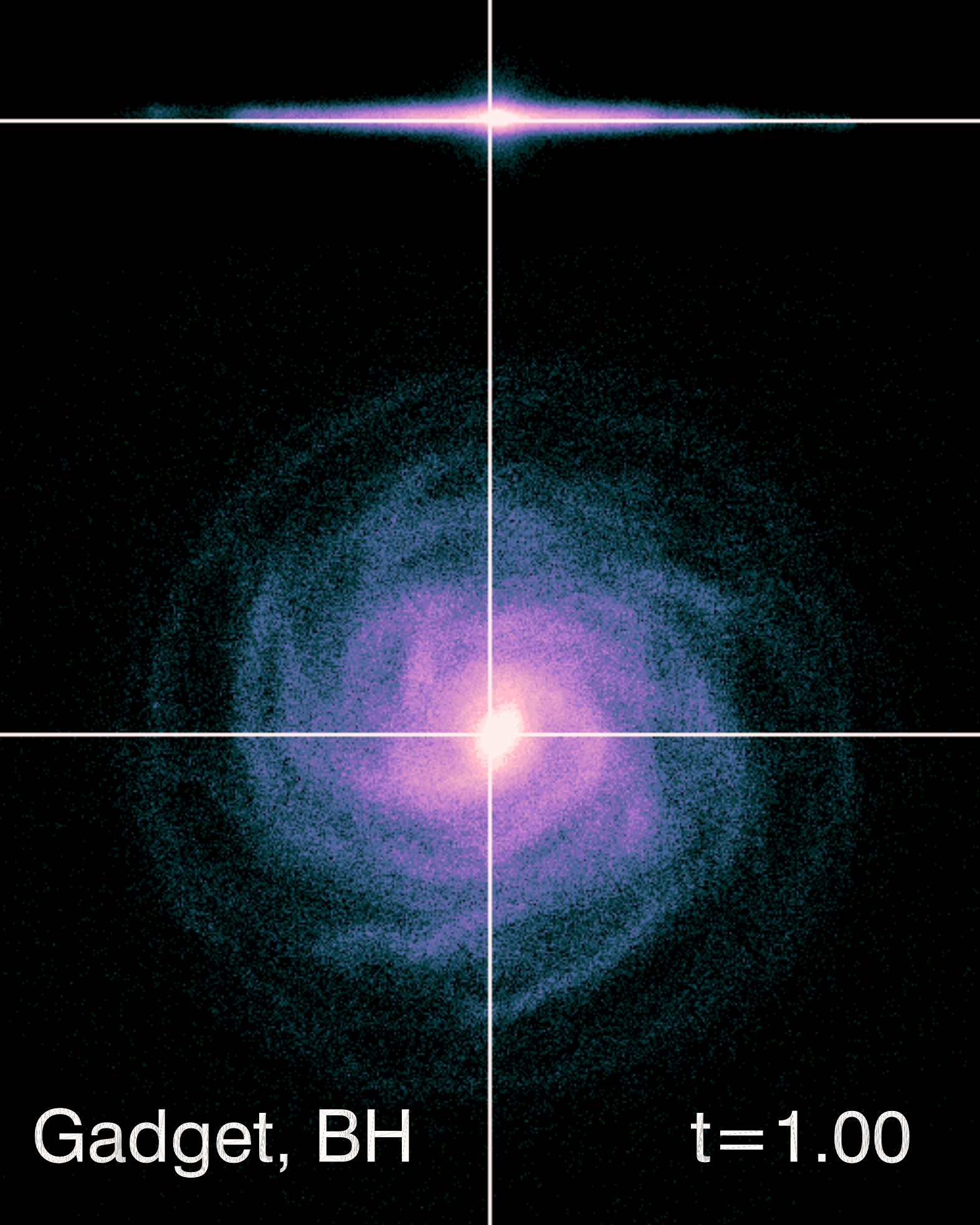}{0.27\textwidth}{}\
              \fig{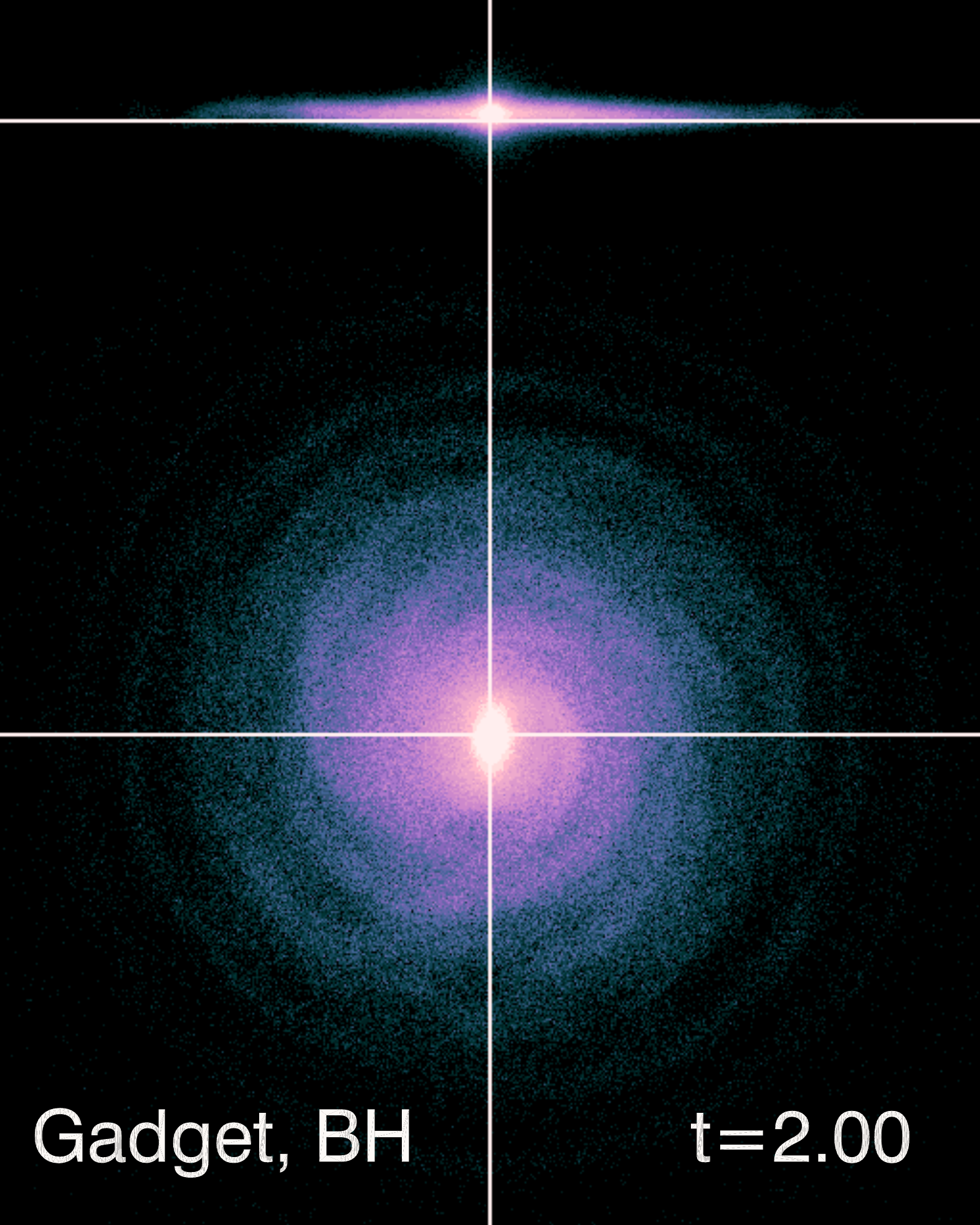}{0.27\textwidth}{}}
\gridline{\fig{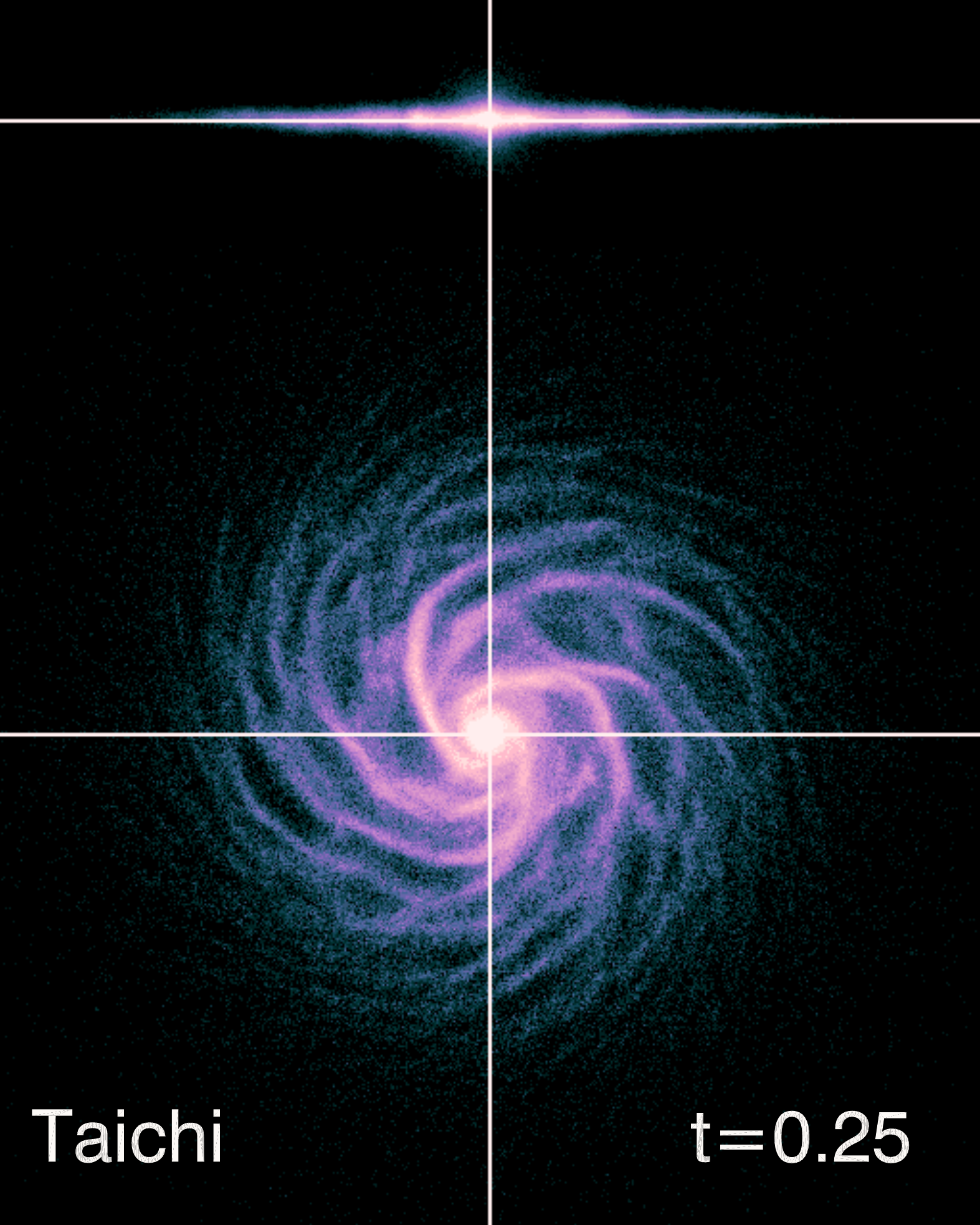}{0.27\textwidth}{}\
             \fig{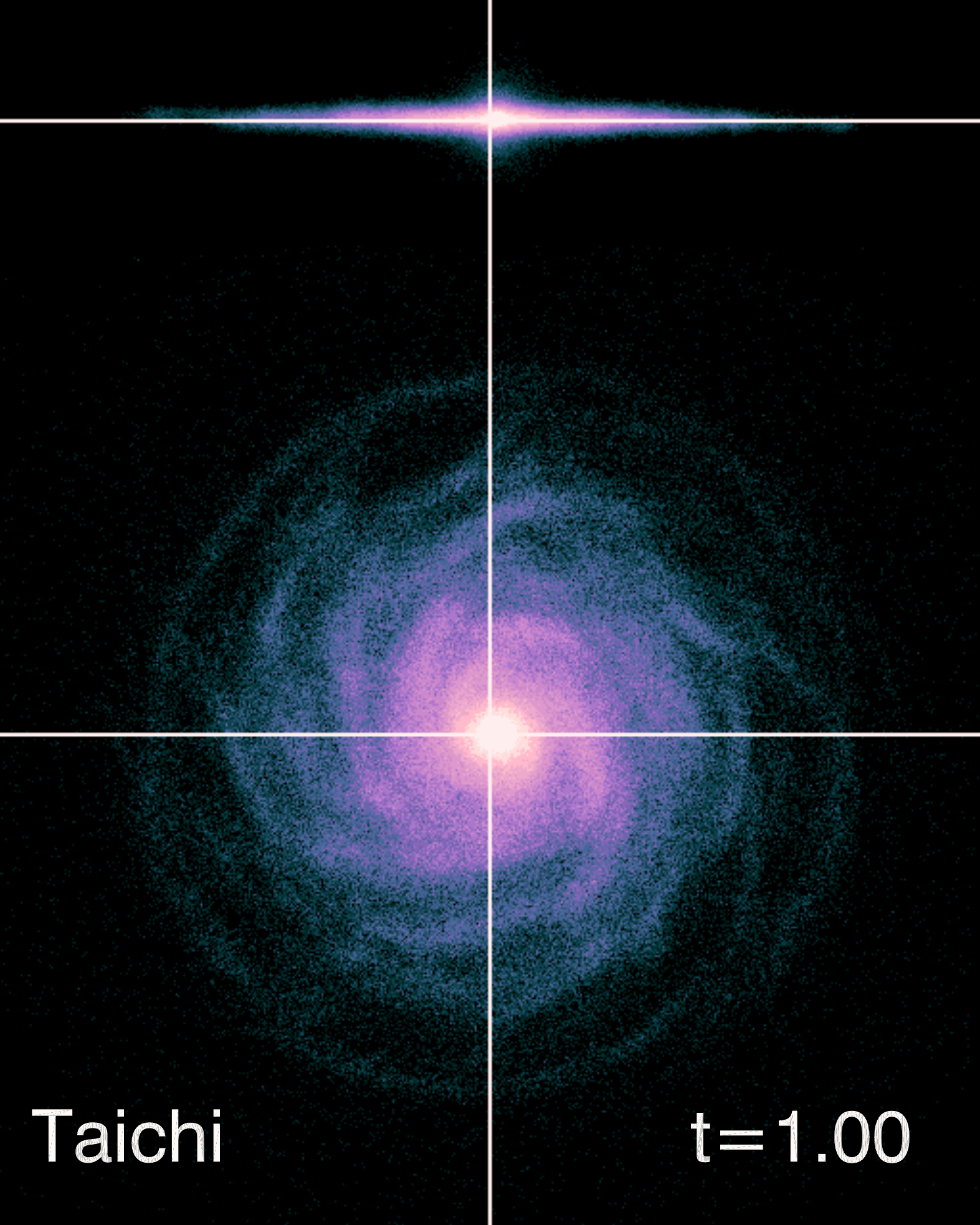}{0.27\textwidth}{}\
             \fig{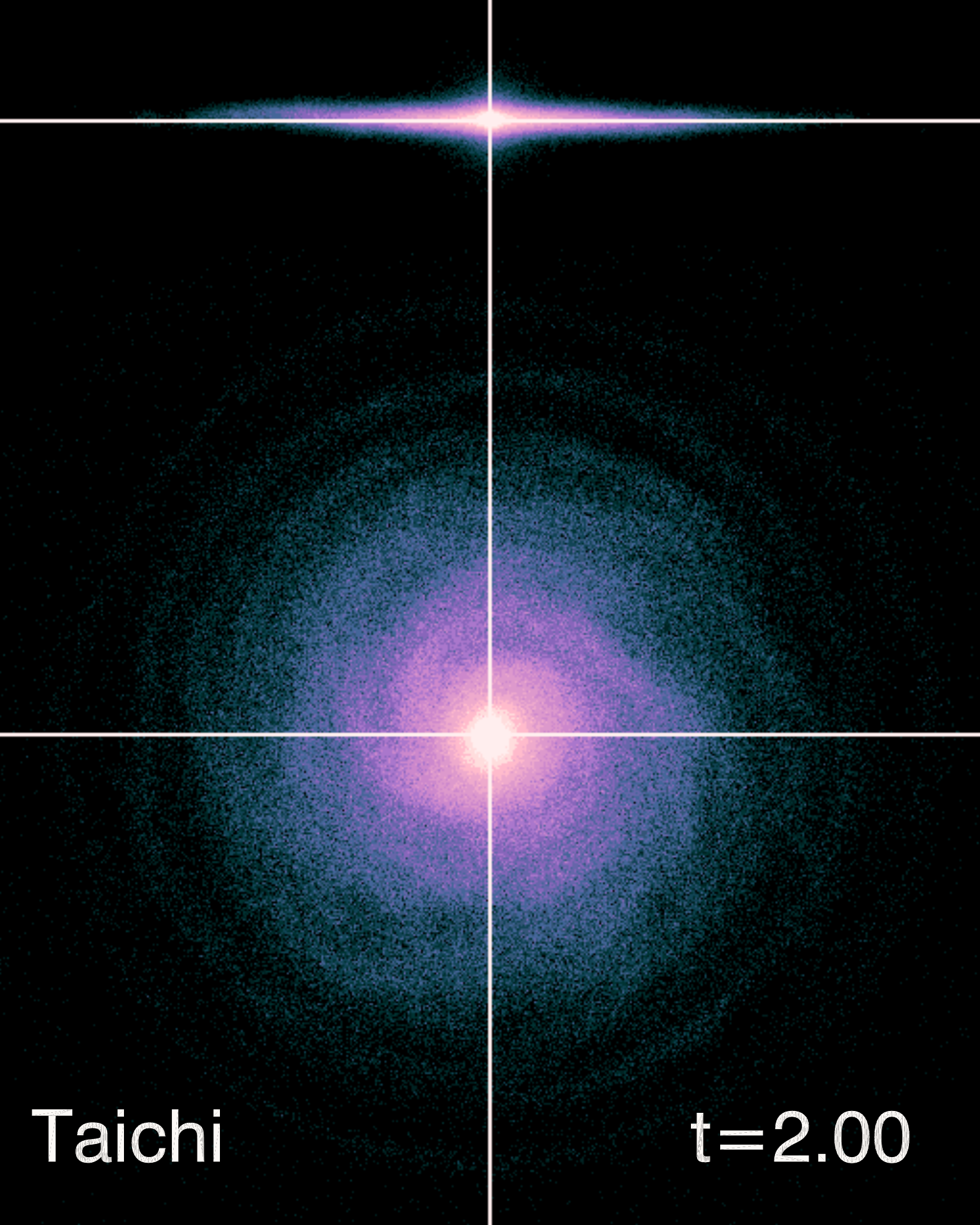}{0.27\textwidth}{}}
\caption{A sequence of images of an isolated disk galaxy simulated with 
{\sc Gadget} and {\sc Taichi}.  We run {\sc Gadget} with the relative opening 
criterion (RC) in this test with $\alpha=2.5\times10^{-3}$. For BH tree method with {\sc Gadget}, 
we use $\theta = 0.45$. 
By $t = 2$, the galaxy has gained a net velocity of 1.4 km/s using {\sc Gadget} with 
the relative opening criterion. This error is visible in the top panels as the galaxy disk 
gradually drift upwards along the $z$-axis. This momentum conservation error is much
smaller with {\sc Gadget} using its BH tree mode with a velocity of 0.1 km/s by $t = 2$.
In contrast, momentum is conserved with the new scheme. An animation of the isolated disk galaxy, 
can be found at \url{https://youtu.be/IaE5OpBhXXk}.
\label{fig:disk}}
\end{figure*}

\subsection{Applications in galactic dynamics}

\begin{figure}[ht!]
\includegraphics[width=\linewidth]{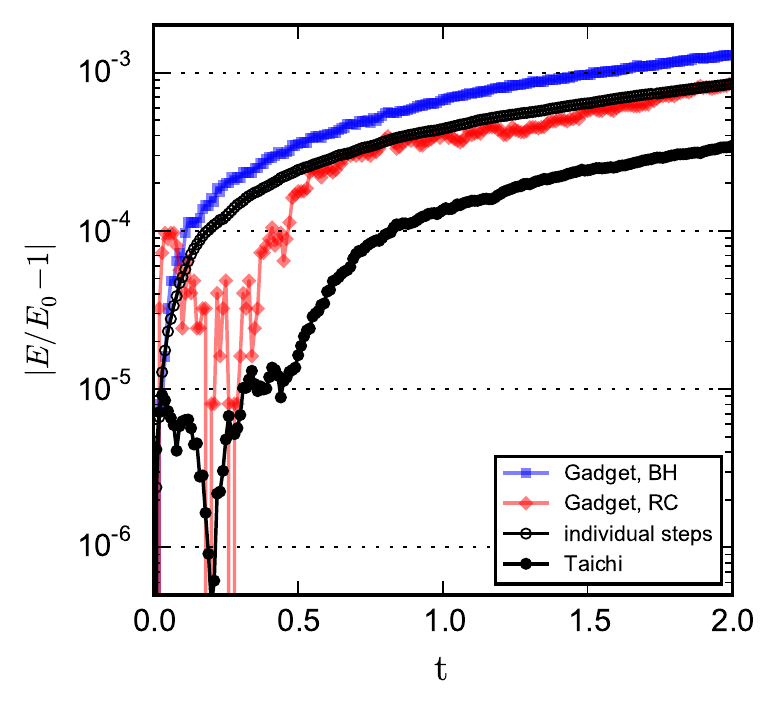}
\caption{\rev{Energy conservation error as a function of time $t$ in the 
isolated disk galaxy test. {\sc Gadget} in its relative opening criterion (RC) and 
in its BH tree mode gives a fractional energy error $\sim10^{-3}$. With the 
traditional individual timesteps, FMM delivers a similar fractional error.
With an additional timestep constraint based
on the total acceleration, {\sc Taichi} is able to reduce the total energy 
conservation error to $3\times10^{-4}$.}
\label{fig:energy_error_evolution}}
\end{figure}

It is quite common to observe isolated galaxy disks drift away from their starting 
point over a long integration time in the literature. This drifting velocity is 
a manifestation of momentum conservation violation in the $N$-body codes. 
Using the method outlined in 
\cite{Springel2005merger}, we set up an exponential stellar disk 
(800,000 particles) with a central bulge (160,000 particles) within a 
dark matter halo (1,000,000 particles) of a Hernquist profile 
\citep{Hernquist1990} with $V_{200} =160$ km/s. We then evolve it with 
{\sc Gadget} and the new scheme. We use $\alpha=2.5\times10^{-3}$ together 
with its relative opening criterion with {\sc Gadget}. 
\rev{The force softening 
length is set to be 15 pc.} By $t = 2$ (1.94 Gyr 
in physical units), the stellar disk drifts upwards along the $z$-axis with 
a net velocity of 1.4 km/s. \rev{If we use the geometric opening angle in BH 
mode with {\sc Gadget} with $\theta=0.45$ instead, the net drifting velocity 
is much smaller, $\sim$~0.1 km/s. As a result, we do not observe any apparent 
drift of the galaxy disk. For the traditional individual timesteps with FMM, 
the net drifting velocity is $\sim$0.02 km/s. 
For comparison, momentum is conserved with the 
new scheme and we do not have any residual drifting velocity at all. }

\rev{It turns out our default timestep function is inadequate for an accurate long term
energy conservation. We find that the fractional 
error in total energy conservation from our isolated disk galaxy simulation is 2.3\% 
by the end of an integration of 4.9 Gyr with {\sc Taichi}. This error is unexpectedly  
large. To fix this problem, 
we have experimented a scheme by imposing a minimum timestep, determined 
by the \textit{total} acceleration due to all particles $\lvert \mathbf{a}\rvert_{\rm tot}$ in 
\begin{equation}
\Delta t_i =  \sqrt{\frac{ \eta_{1} \epsilon_i}{\lvert \mathbf{a}\rvert_i}}.
\label{eq:stepfunction1}
\end{equation}
During the simulation, $\lvert \mathbf{a}\rvert_{\rm tot}$ can be obtained in 
a kickSF2S step when all particles are passed to the Poisson solver, i.e., when 
the bottom rung is evaluated. With this additional timestep constraint, the total 
energy conservation with {\sc Taichi} performs very well when compared 
to {\sc Gadget}. The fractional energy conservation error is about 
$3\times10^{-4}$, slightly better than the result obtained with {\sc Gadget} in its relative 
opening criterion mode. Interestingly, while momentum conservation error is 
quite small with {\sc Gadget} in its BH tree mode, the energy 
conservation error is quite large, reaching to $10^{-3}$ by $t = 2$. 
Therefore, it appears tricky to achieve good momentum conservation 
and energy conservation simultaneously with {\sc Gadget}.}

\rev{In this test, {\sc Taichi} finishes with a speed comparable to 
{\sc Gadget} using the same CPU cores. However, given the 
similar computational cost, {\sc Taichi} offers much higher accuracies. 
{\sc Taichi} is also 40\% faster than FMM code using the traditional 
individual timesteps.}  
  
\subsection{Applications in cosmological simulations}

The new $N$-body scheme can be readily extended to cosmological 
simulations. The high accuracy of FMM can also 
be beneficial where the mass distribution is nearly uniform at high redshift.
Periodic boundary conditions can be treated either with Ewald summation 
\citep{Hernquist1991} as in \cite{Potter2017} or with a renormalization 
procedure using the same cell--cell interactions within FMM \citep{Berman1994}, 
which {\sc ExaFMM} has already implemented . 

The cosmological integration is straightforward following the 
{\sc Gadget} code paper by \cite{Springel2005}. 
We use the comoving coordinate $\mathbf{x}$ and its
canonical momentum $\mathbf{p} (\equiv a^2 m \mathbf{x}^2)$ with 
the scale factor $a$. The cosmological factors in the drift operator
\begin{equation}
\mathbf{x} \rightarrow \mathbf{x}  + \frac{\mathbf{p}}{m} \int_{t}^{t+\Delta t} \frac{dt}{a^2},
\end{equation}
and in the kick operator
\begin{equation}
\mathbf{p} \rightarrow \mathbf{p} + (-\nabla \Phi) \int_{t}^{t+\Delta t} \frac{dt}{a},
\end{equation}
are stored in lookup tables as a function of $a$. 

\begin{figure*}
\gridline{\fig{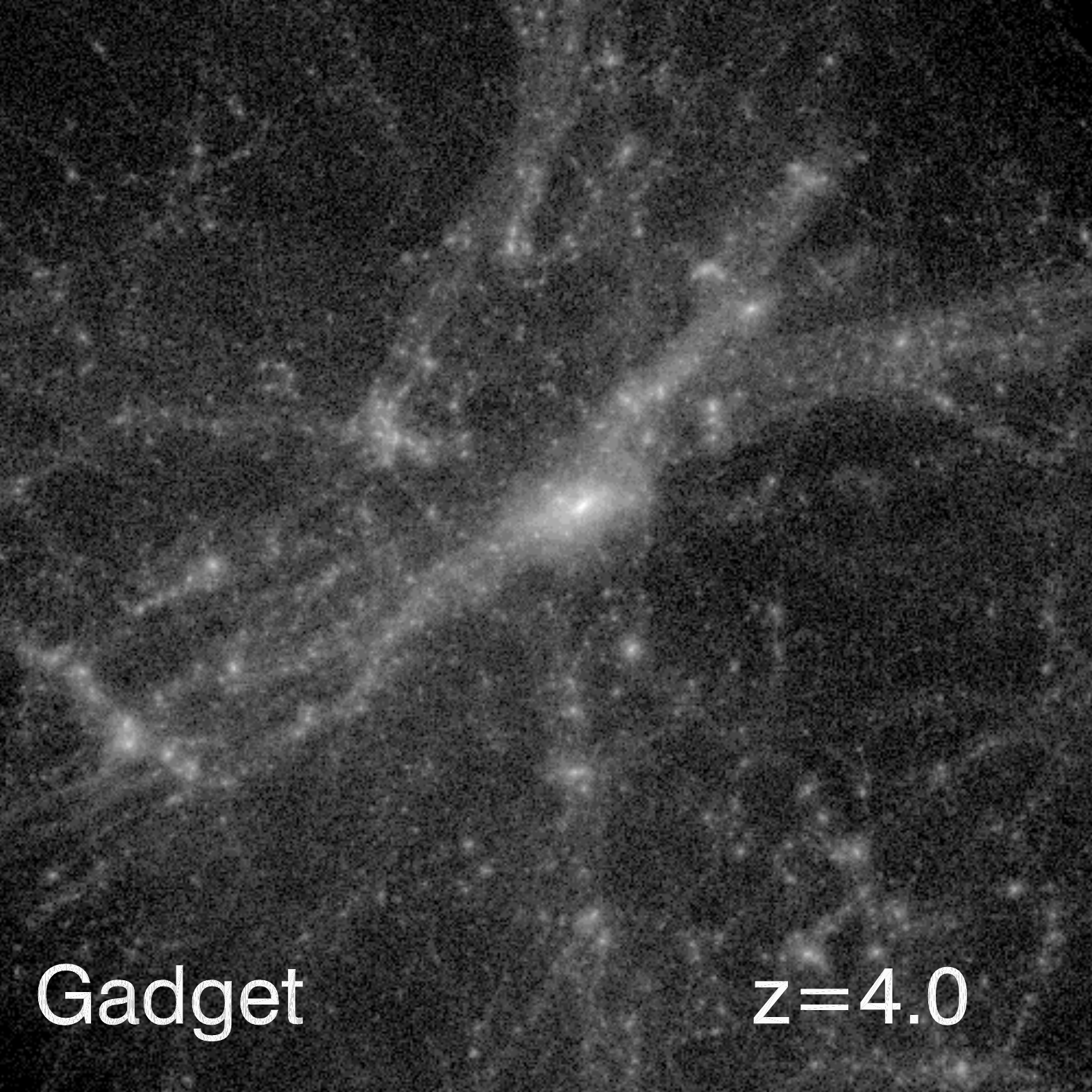}{0.25\textwidth}{}
              \fig{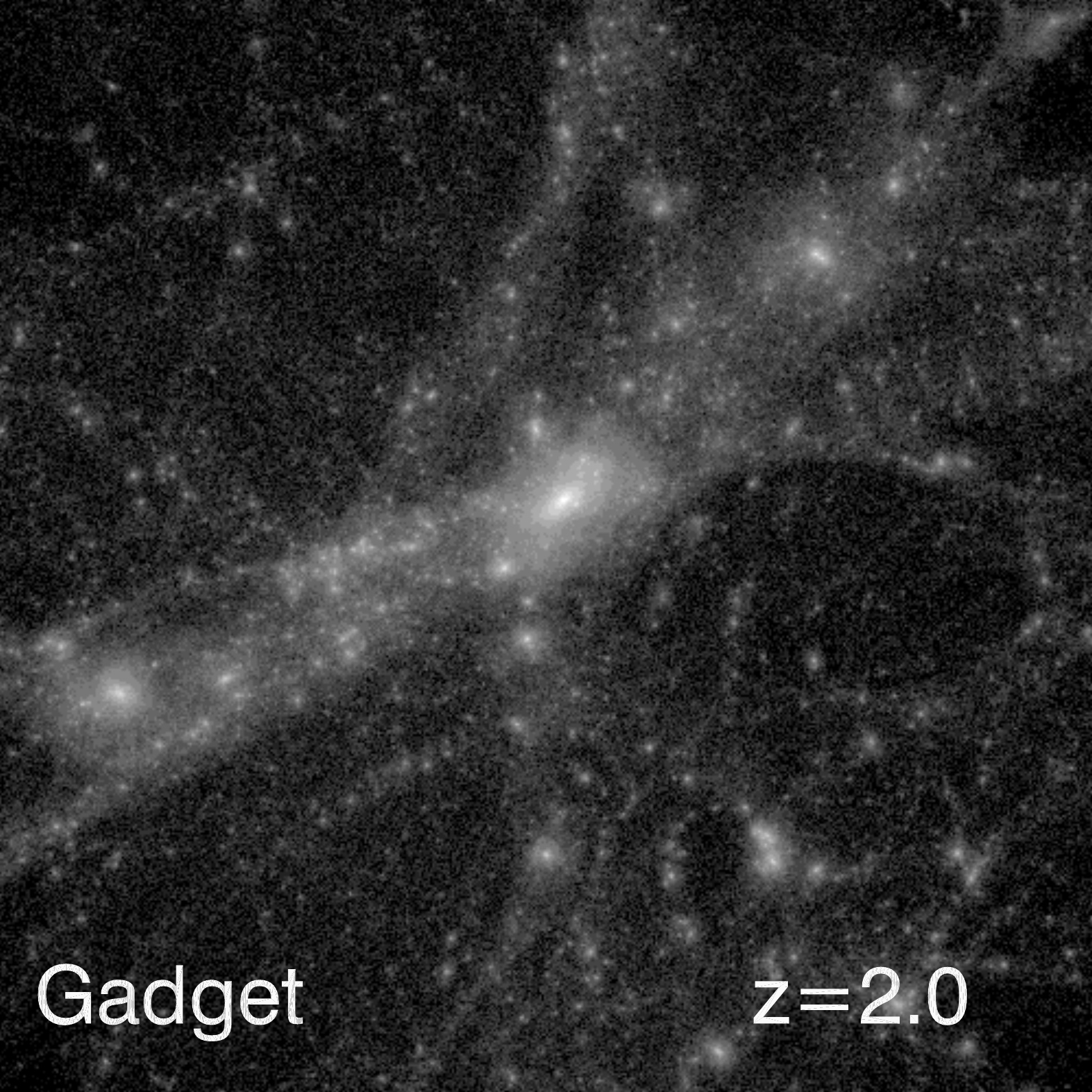}{0.25\textwidth}{}
              \fig{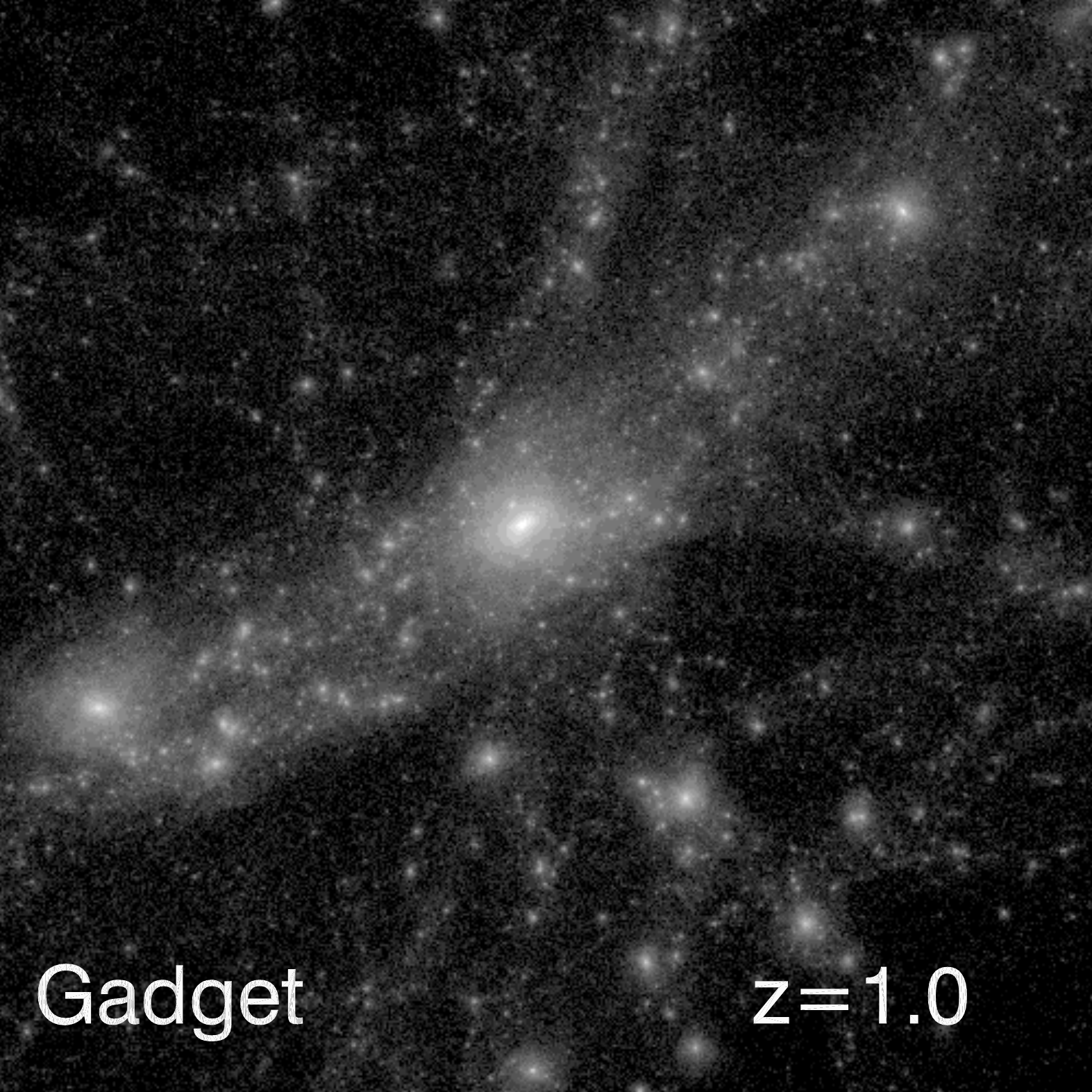}{0.25\textwidth}{}
              \fig{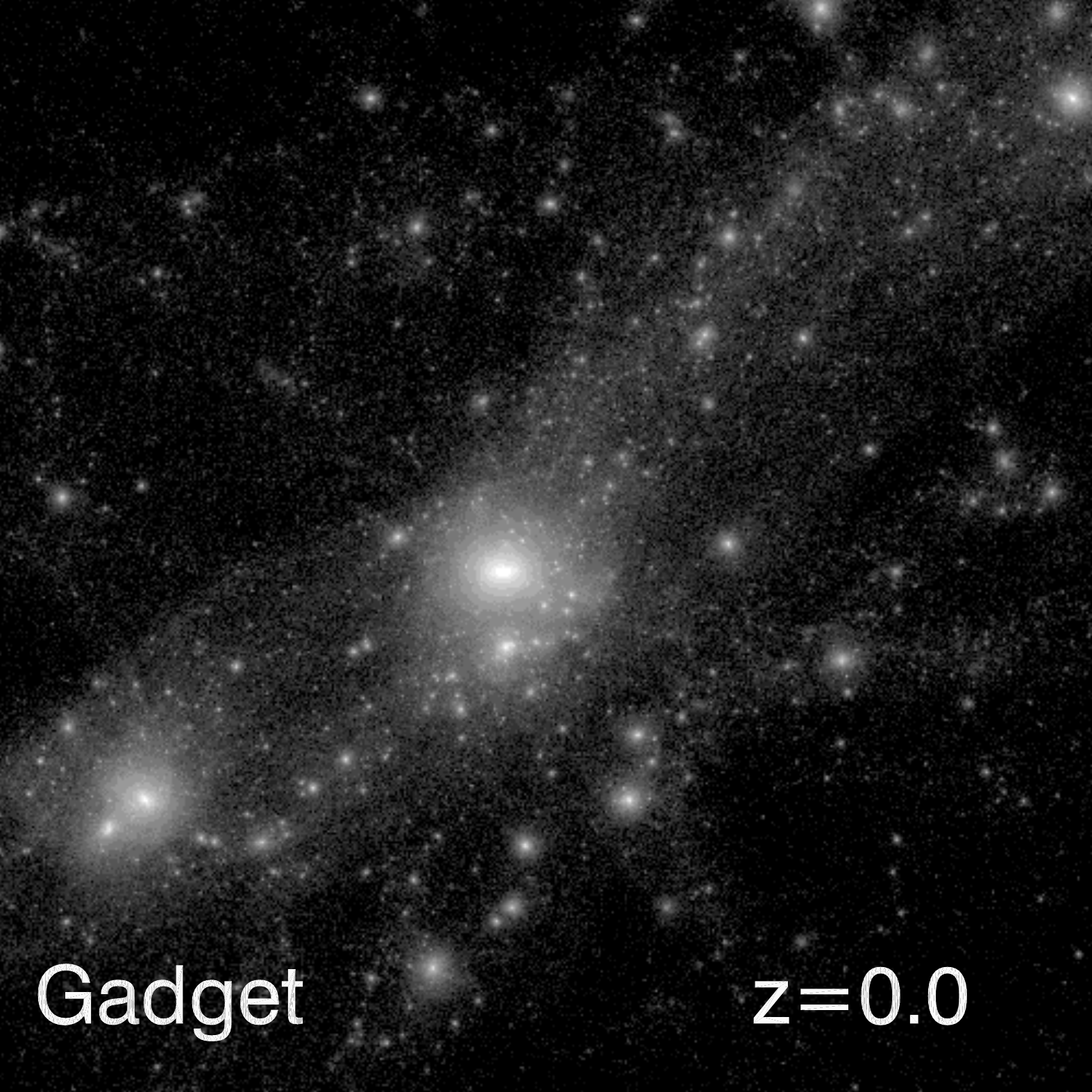}{0.25\textwidth}{}}
\vspace{-0.85cm}
\gridline{\fig{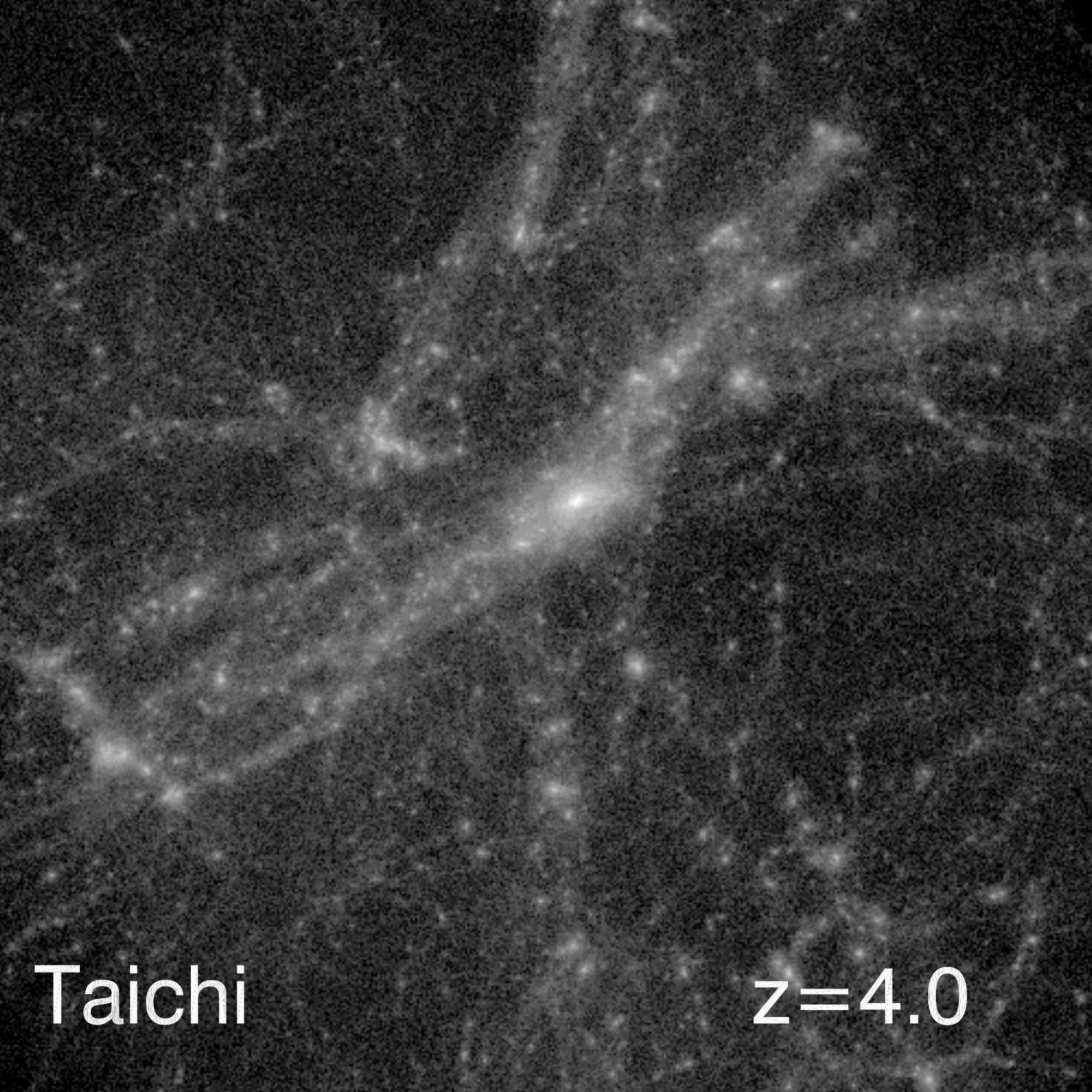}{0.25\textwidth}{}
              \fig{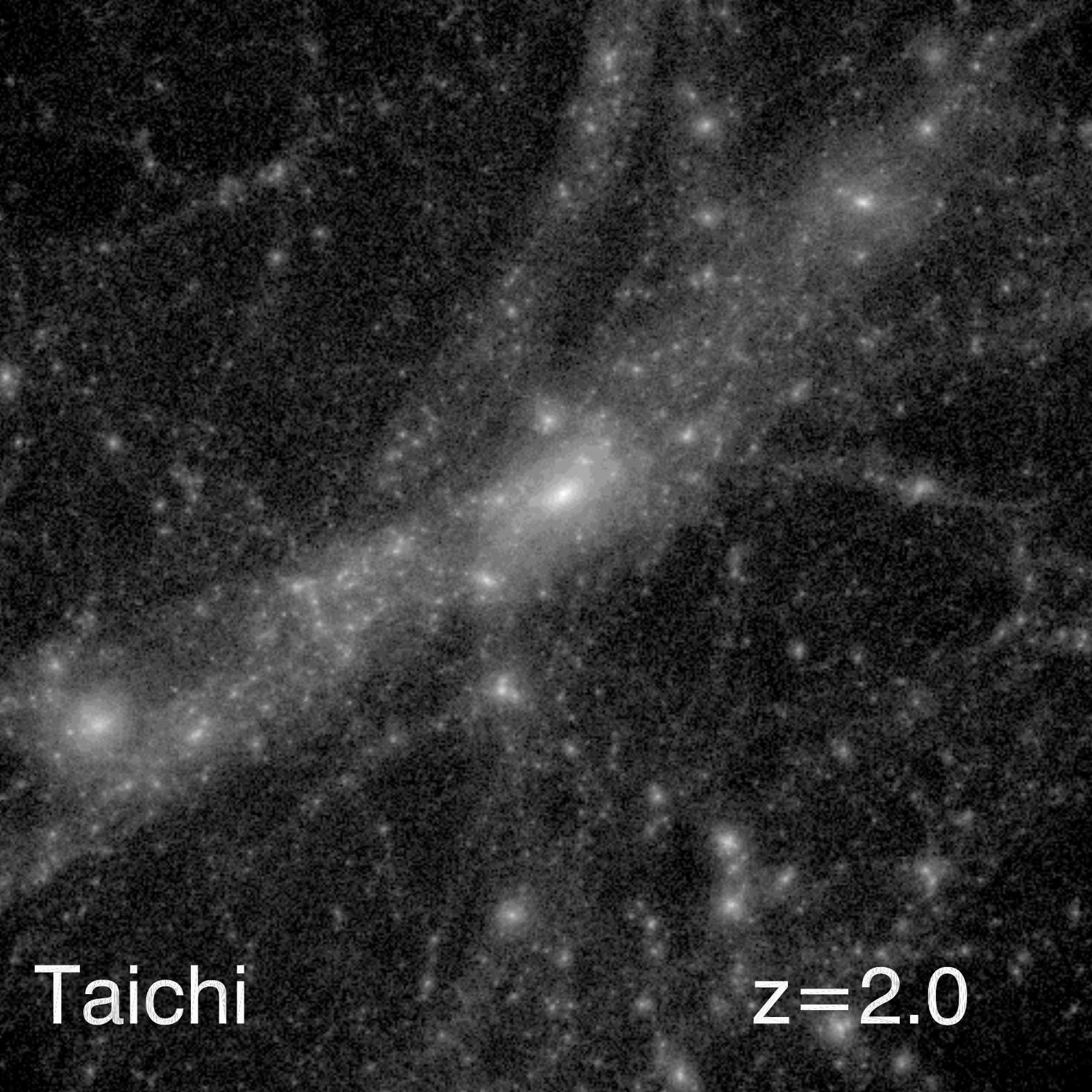}{0.25\textwidth}{}
              \fig{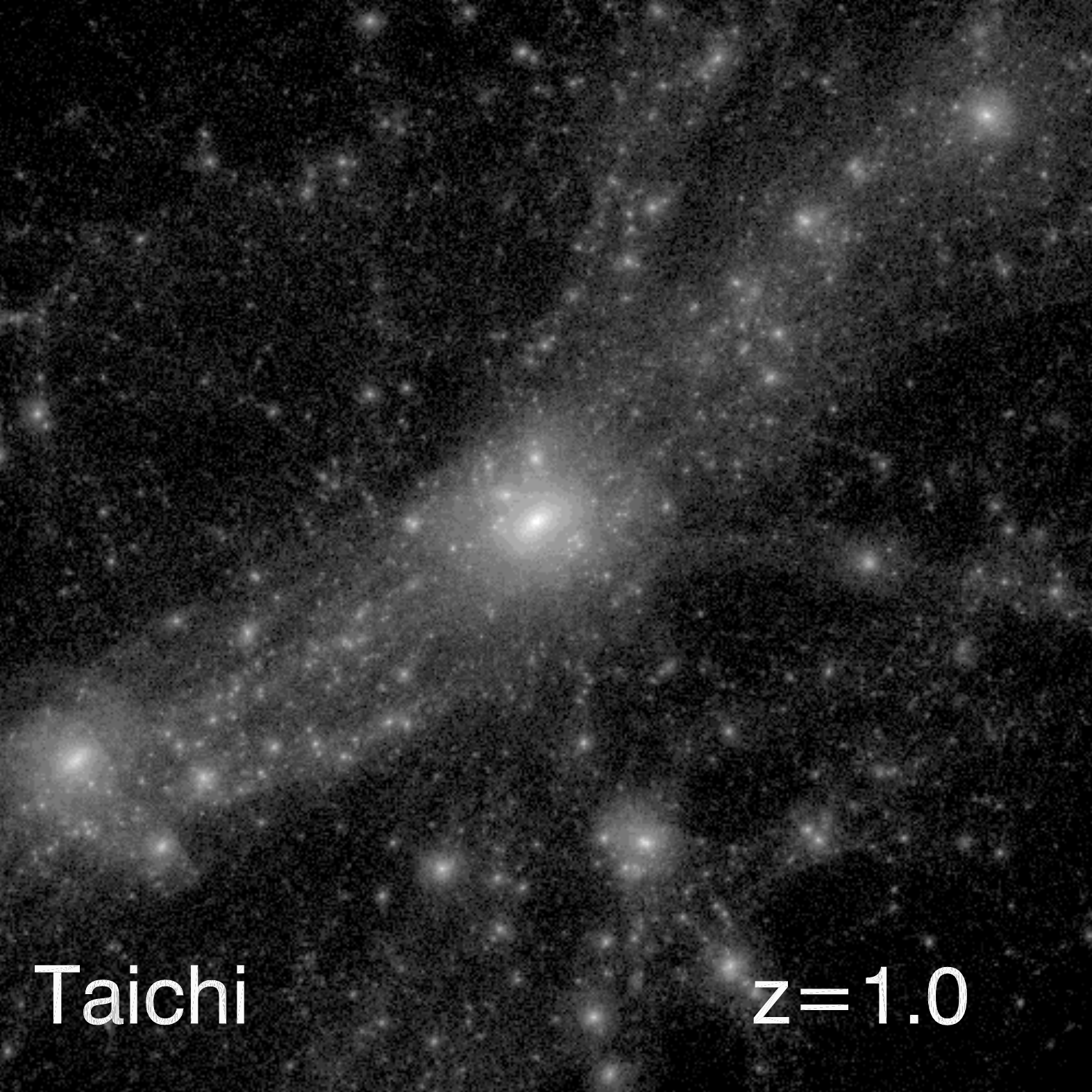}{0.25\textwidth}{}
              \fig{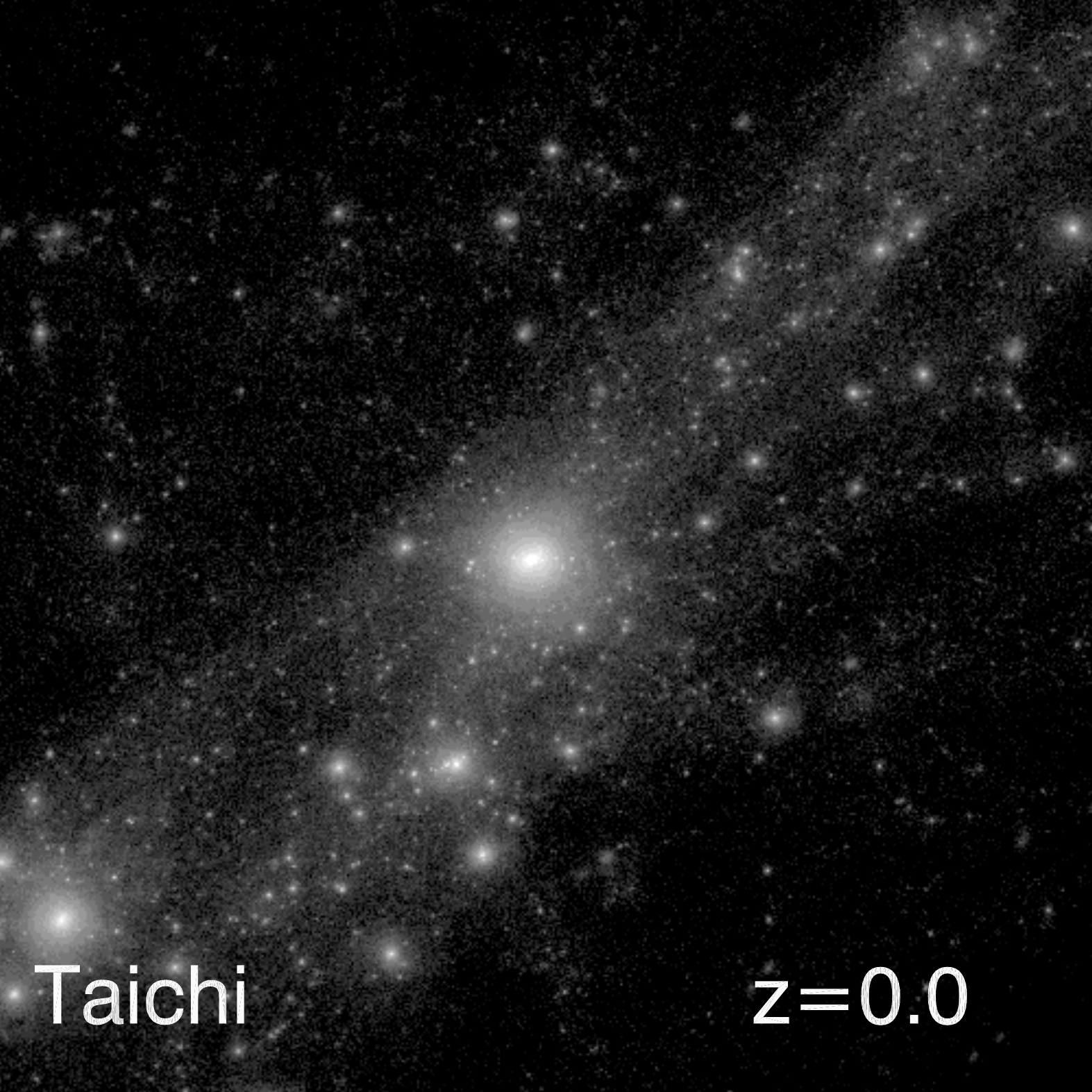}{0.25\textwidth}{}}  
              
\caption{Dark matter distributions of Aq-C-5 halo simulated with {\sc Gadget} (in its TreePM mode)
and {\sc Taichi} 
at redshift $z = 4$, 2, 1 and 0.
\label{fig:aquila}}
\end{figure*}

We integrate the initial condition of a Milky Way-sized halo in its a dark matter only 
version\footnote{ICs are available at 
\url{http://www.aip.de/People/cscannapieco/aquila/ICs/Gadget/Aq-C-5/}.}. This halo is
Aq-C-5 in the Aquarius project \citep{Springel2008} and its hydrodynamic version is 
used for the Aquila comparison project \citep{Scannapieco2012}.
Figure~\ref{fig:aquila} shows the dark matter distribution in its projected mass density 
at redshift $z = 4$, 2, 1, and 0. The top panels show the formation of the dark matter halo
with {\sc Gadget} (in its TreePM mode) and the lower with {\sc Taichi}. 
Qualitatively, dark matter filaments, 
halos, and halo substructures are very similar between {\sc Gadget} and {\sc Taichi}.
\rev{The raw speed of {\sc Taichi} is much slower than {\sc Gadget}
due to the fact {\sc Gadget} only uses the tree method to calculate the short 
range force. We note that the speed comparison between the two codes
in this case is unfair because the force accuracies are not on the same level. 
If we run {\sc Gadget} with only the tree method, it is much slower than 
{\sc Taichi}, as expected.} \rev{In principle,} particle mesh method could 
also be integrated with FMM as a long-range force solver as in TreePM codes.  
However, it is not clear whether FFT is necessary after all because FFT 
complicates the error analysis \citep[see also][]{Ishiyama2009, Schneider2016}. 
\rev{The benefits with a hybrid FFT+FMM solver are unwarranted. 
The performance of FFT has become a major bottleneck in many large scale
applications. On the other hand, \cite{Yokota2014} have argued FMM has
very attractive communication complexities. 
Also, a more efficient tree walk and further optimizations in the FMM solver
can considerably speed up {\sc Taichi}.}

\section{Discussions} \label{sec:discussion}
\subsection{Major shortcomings of the new $N$-body scheme}
\label{subsec:stepsize}

\rev{As shown in the isolated galaxy disk test, it is also necessary to impose an
additional constraint to get excellent energy conservation. This highlights 
one primary shortcomings of this new scheme: the full acceleration on 
each particle can only be computed when \emph{all} the particles are 
passed in the Poisson solver, which can be inadequate for the highly
dynamical regions in some cases.}

\rev{One possible remediation to this shortcoming is to use a higher 
order integrator, such as a forward step-only fourth-order integrator 
proposed by \cite{Chin2005}. This integrator is a promising alternative 
to the traditional second-order KDK integrator. 
Although the approximation of the force gradient $\nabla \mathbf{a}_i$ is 
lower than the acceleration itself, FMM should give a reasonably 
accurate measure of $\nabla \mathbf{a}_i$ with a high expansion order. 
Our preliminary tests show that the fourth order integrator by \cite{Chin2005} 
gives at least two orders of magnitude smaller energy error even with variable 
timesteps than the second order KDK integrator given the same cost.}

\rev{In the absence of a totally satisfactory treatment of timestep function, 
we would recommend to use some monitoring of total energy, 
momentum, and angular momentum during the course 
of a simulation as suggested by \cite{Dehnen2011}. As demonstrated 
by the isolated disk galaxy, {\sc Taichi} can deliver excellent momentum
and total energy conservation in its current form. }

\begin{figure}[ht!]
\includegraphics[width=\linewidth]{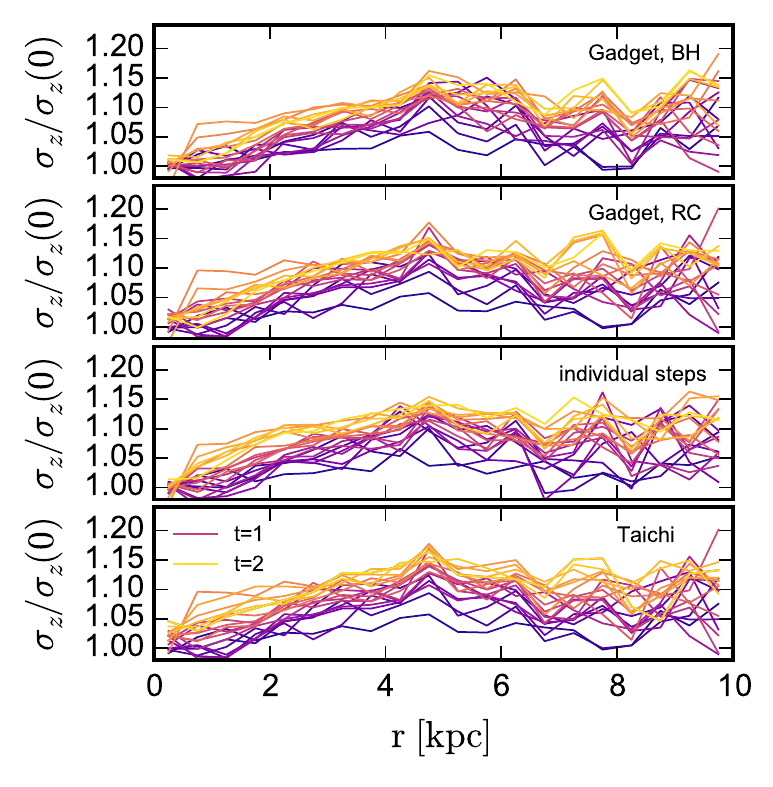}
\caption{\rev{Time evolution of the vertical velocity dispersion profiles 
of the stellar disk in our isolated galaxy test. Each 
profile is normalized by the vertical velocity dispersion at $t = 0$. }
\label{fig:disk_velz_disp_evolution}}
\end{figure}

\rev{We should also point out another potential shortcoming with the
new $N$-body scheme. In the collision-less system, particles 
should follow the trajectories determined by the smooth potential. 
However, the hierarchical splitting introduces additional perturbations
on top of the existing potential fluctuations \citep[e.g.,][]{Hernquist1993}. 
We can use the vertical thickening of the initially thin disks as a useful 
diagnose of collisional heating \citep{Sellwood2013}. In 
Figure~\ref{fig:disk_velz_disp_evolution}, we show the evolution of 
vertical velocity dispersion of the stellar disk in our previous test. 
The vertical velocity dispersion gradually but slowly increases as 
a function of time. And the time evolution of velocity dispersion profiles
obtained with our four simulations is also consistent with each other.}

\rev{We have also compared the evolution of a Hernquist halo using {\sc Taichi}
and {\sc Gadget} and found the evolution of the central density cusp 
are very similar to each other. It indicates that shot noise 
\citep{Hernquist1993} is likely to dominate over the additional 
perturbations from the hierarchical splitting. 
It is safe to conclude the new scheme is as good as 
the existing ones treating collision-less dynamics}.

\subsection{Roads towards fast and accurate $N$-body schemes 
for collision-less and collisional dynamics}
As an approximate Poisson solver, FMM is suitable for  
a rigorous error analysis following the classic text of \cite{Kellogg} and the 
error analysis in the seminal work of \cite{Greengard1987}. 
When coupled with the HHS time integrator, we have 
demonstrated that it is possible to construct a practical $N$-body scheme 
with individual timesteps that (1) retains the \ON\ algebraic complexity of FMM,
(2) \rev{conserves linear momentum accurate to the round-off errors}.
Both properties are highly desired in a broad range of astrophysical 
applications as well as in other fields such as electrostatic plasma 
simulations and molecular dynamics. 

Currently, our understandings of dark matter distributions have 
almost exclusively relied on numerical simulations. 
It has been shown recently that halo density profiles and shapes are 
affected by the dynamics of baryons as well \citep[e.g.,][]{Pontzen2014,
Chan2015}. The distribution of dark 
matter essentially reflects the shape of individual orbit \citep{More2015, Zhu2017}.
It is therefore sensible to invest in a more accurate $N$-body scheme 
to compare with the current codes before any complex baryonic dynamics are 
included. Force accuracies in the current $N-$body codes are already hinted in 
a recent study of matter clustering by \cite{Schneider2016} where three different 
codes show discrepancies bigger than one per cent level in matter power spectra. 

We could also improve the accuracy of the proposed new scheme with some 
time-symmetric stepping formulation \cite{Dehnen2017}. The improvements 
are within a single and consistent framework. After finishing the first draft, 
we learned that a combination of FMM and HHS has also already been 
incorporated into the latest version of {\sc Gadget} for a while
(Volker Springel, private communication;
Springel et.~al, in prep). FMM with HHS integrator has a potential to be 
a highly accurate code for computational cosmology.

On the computational side, the \ON\ algebraic complexity and 
the low communication complexity makes FMM petascale friendly 
\citep[e.g.,][]{Lashuk2012, Yokota2012}. The expensive direct
summation part of FMM can be accelerated with either GPU or
AVX instructions. Also, multipole expansion in the Cartesian 
coordinates runs much faster than in spherical coordinates 
when $p$ is small \citep{Dehnen2000, Yokota2012b}. We can then 
customize this scheme with a low order expansion for collision-less 
simulations accordingly due to their Monte Carlo nature.

This study focuses on collision-less systems such as galaxies
and dark matter halos. The HHS time integrator by \cite{Pelupessy2012} is 
proposed in the context of collisional dynamics, though a more efficient 
splitting has been proposed by \cite{Janes2014}. Another possible follow-up
development of this study is to incorporate an error-controlling MAC 
\citep{Dehnen2014} to make the Poisson solver accurate enough for 
collisional systems. It is quite reasonable to expect that many currently 
prohibiting problems in collisional dynamics can benefit from the 
proposed new $N$-body scheme. 

\section{Summary} \label{sec:summary}
In this paper, we discuss the origins of momentum conserving error in 
the previous $N$-body codes and propose an exact momentum-conserving
scheme with a Poisson solver based on FMM and a time integrator based on HHS. 
We implement the new $N$-body scheme in the {\sc Taichi} code. Using cold 
collapse test,  we show the combination of these two momentum conserving 
components also improves both angular momentum and energy conservations. 
We then test {\sc Taichi} in an isolated disk galaxy evolution and a cosmological 
structure formation simulation. Our main findings are

\begin{itemize}
\item Traditional $N$-body codes violate Newton's third law due to force 
asymmetries in space from approximate Poisson solvers and asymmetries
in time when individual timesteps are employed. 

\item The new $N$-code scheme, on the other hand,  
is momentum conserving by construction. The momentum-conserving 
nature of the new scheme is verified with our implementation in the 
{\sc Taichi} code. 

\item The \ON\ complexity of FMM can be retained in the new scheme
as the Hamiltonian splitting continuously reduces the problem size for more
dynamic part of the system.

\item The new $N$-code scheme can be readily extended to cosmological
simulations, as we have shown with a zoom-in simulation of a Milky-Way
sized halo. 

\end{itemize}

\rev{The main shortcoming of the new $N$-code scheme is the lack of 
an accurate and reliable timestep function. Future work also include a 
higher order integrator, multipole expansions in the Cartesian coordinates,  
and an efficient tree walk with error controls. There is also large room to 
improve the code efficiency of with various optimizations. Lastly, 
we would like to emphasize this work does not in any way imply 
the simulations done by the current codes are inaccurate/incorrect. 
In many applications of collision-less systems, a modest accuracy 
is already sufficient. When high accuracies are required, new schemes
such as the one proposed in this study can have advantages over the 
traditional schemes.}


\acknowledgments
We express our gratitude towards the referee,  Walter Dehnen, for constructive and insightful 
reports which helped us to improve this paper on the most fundamental level. 
It is a pleasure to thank Yuexing Li, Lars Hernquist, and Volker Springel for their 
helpful discussions and comments. 
Special thanks to Rio Yokota and F.I.~Pelupessy for making their codes public. 
This work was supported by NSF grants AST-0965694, AST-1009867, AST-1412719, 
and MRI-1626251. The numerical computations and data analysis in this paper 
have been carried out on the CyberLAMP cluster supported by MRI-1626251, 
operated and maintained by the Institute for CyberScience at the Pennsylvania State 
University, as well as  the Odyssey cluster supported by the FAS Division of Science, 
Research Computing Group at Harvard University. The Institute for Gravitation and 
the Cosmos is supported by the Eberly College of Science and the Office of the 
Senior Vice President for Research at the Pennsylvania State University.

%\bibliography{ref} 

\clearpage\end{CJK}
\end{document}